\documentclass{sig-alternate-2013}

\newfont{\mycrnotice}{ptmr8t at 7pt}
\newfont{\myconfname}{ptmri8t at 7pt}
\let\crnotice\mycrnotice%
\let\confname\myconfname%

\usepackage{balance}  
\usepackage{algorithmic}
\usepackage{algorithm}
\usepackage{multirow}
\usepackage{graphicx}
\usepackage{times}
\usepackage{url}
\usepackage{amsmath}
\usepackage{amssymb}
\usepackage{amsfonts}
\usepackage{xspace}
\usepackage{url}
\usepackage{./style/tweaklist}
\usepackage[show]{style/chato-notes}
\usepackage{color,soul}

\newcommand{\IGNORE}[1]{}

\newcommand{\commentedtext}[1]{}

\newtheorem{example}{Example}

\newcommand{\eat}[1]{{}}

\newcommand{\algoname}{\textsc{TOPKS-ASYT}}

\newcommand{\indexname}{\textsc{CT-IL}}

\permission{Permission to make digital or hard copies of all or part of this work for personal or classroom use is granted without fee provided that copies are not made or distributed for profit or commercial advantage and that copies bear this notice and the full citation on the first page. Copyrights for components of this work owned by others than ACM must be honored. Abstracting with credit is permitted. To copy otherwise, or republish, to post on servers or to redistribute to lists, requires prior specific permission and/or a fee. Request permissions from Permissions@acm.org.}
\conferenceinfo{CIKM'15,}{October 19--23, 2015, Melbourne, Australia.}
\copyrightetc{\copyright~2015 ACM. ISBN \the\acmcopyr}
\crdata{978-1-4503-3794-6/15/10\ ...\$15.00.\\
DOI: http://dx.doi.org/10.1145/XXX.XXXXXXX}

\begin{document}

%

\title{A Network-Aware Approach for Searching \\As-You-Type  in Social Media}
\subtitle{Extended Version}


\numberofauthors{3}
\author{
\alignauthor
Paul Lagr\'{e}e \\
      \affaddr{Inria Saclay and Universit\'e Paris-Sud}\\
      \affaddr{Orsay, France}\\
       \email{paul.lagree@u-psud.fr}
\alignauthor
Bogdan Cautis\\
       \affaddr{Inria Saclay and Universit\'e Paris-Sud}\\
       \affaddr{Orsay, France}\\
       \email{bogdan.cautis@u-psud.fr}
\alignauthor
Hossein Vahabi  \\
       \affaddr{Yahoo Labs}\\
       \affaddr{Barcelona, Spain}\\
       \email{puya@yahoo-inc.com}
}

\maketitle \sloppy

\begin{abstract}
We present in this paper a novel approach for as-you-type top-$k$ keyword search over social media. We adopt a natural ``network-aware'' interpretation for information relevance, by which information produced by users who are closer to the seeker is considered more relevant. In practice, this query model poses new challenges for effectiveness and efficiency in online search, even when a complete query is given as input in one keystroke. This is mainly because it requires a joint exploration of the social space and classic IR indexes such as inverted lists. 
 We describe a memory-efficient and incremental prefix-based retrieval algorithm, which also  exhibits an anytime behavior, allowing to  output the most likely answer within any chosen running-time limit. 
  We evaluate it through extensive
 experiments for several applications and search scenarios, including searching for  posts in micro-blogging (Twitter and Tumblr), as well as  searching for businesses based on  reviews in Yelp.  They show that our solution is effective in answering real-time as-you-type searches over social media. 

\end{abstract}




\section{Introduction}
\label{sec:introduction}

Information access on the Web, and in particular on the social Web, is,  by and large,  based on top-$k$ keyword search. While we  witnessed significant improvements  on how to answer keyword queries on the Web in the most effective way (e.g., by exploiting the Web structure, user and contextual models, user feedback, semantics, etc), answering information needs in social applications requires often a significant departure from socially-agnostic approaches, which generally assume that the data being queried is decoupled from the users querying it.   The rationale is that  social links can be exploited in order to obtain more relevant results, valid not only with respect to the queried keywords but also with respect to the social context of the user who issued them.

While progress has been made in 
recent years to support this novel,  social and  \emph{network-aware},  query paradigm --  especially towards efficiency and scalability --  more remains to be done in order
to address  information needs in real applications. In particular, providing the most \emph{accurate} answers while the user is typing her query, \emph{almost instantaneously},   can be extremely  beneficial, in order to enhance  the user experience and to guide the retrieval process.

 In this paper, we  adapt and extend to the as-you-type search scenario  -- one by now  supported in most search applications, including  Web search  --  existing algorithms for top-$k$ retrieval over social data. Our solution, called \algoname\ (for TOP-$k$ Social-aware search AS-You-Type),  builds on the generic network-aware search approach of~\cite{Maniu13-1,Schenkel08} and deals with  three systemic changes:  
\begin{enumerate}
\vspace{-1mm}
\item \textit{Prefix matching:}  answers must be  computed following a query interpretation by which the last term in the query sequence can match tag / keyword  prefixes.
\vspace{-1.5mm}
\item \textit{Incremental computation:}   answers must  be  computed \emph{incrementally}, instead of  starting a computation from scratch. For a query representing a sequence of terms (keywords) $Q=[t_1, \dots ,t_r]$, we can follow an approach that exploits  what has already been computed in the query session so far, i.e., for the query  
 $Q'=[t_1, \dots ,t_{r-1}, t_r']$, with $t_r'$ being a one character shorter prefix of the term $t_r$. 
 \vspace{-1.5mm}
 \item \textit{Anytime output:} answers, albeit approximate,  must be ready to be outputted at any time, and in particular after any given time lapse (e.g., $50-100ms$ is generally accepted as a reasonable latency for as-you-type search).   
\end{enumerate}

We consider a generic setting common to a plethora of social applications, where users produce unstructured content  (keywords)  in relation to items, an activity we simply refer to as \emph{social tagging}.  More precisely, our core application data can be modelled as follows: (i) users form a social network, which may represent relationships such as 
similarity, friendship, following, etc, (ii) items from a public pool of items (e.g., posts, tweets,
videos, URLs, news, or even users) are ``tagged'' by users with keywords, through various interactions and data publishing scenarios, and (iii) users search for some $k$ most relevant items by keywords. 

We devise a novel index structure for \algoname, denoted \indexname, which is a combination of tries and inverted lists. While basic trie structures have been used in as-you-type search scenarios in the literature (e.g., see~\cite{Li:2012} and the references therein), ranked access over inverted lists requires an approach that performs  ranked completion more efficiently.  Therefore, we rely on a trie structure tailored  for the problem at hand, offering a good space-time tradeoff, namely the \emph{completion trie }of~\cite{Hsu:2013}, which is an adaptation of the well-known Patricia trie using priority queues. This data structure is used as the access layer over the inverted lists, allowing us to read in sorted order of relevance the possible keyword completions and the items for which they occur.  Importantly, we use  the completion trie not only as an index component over the database, 
 but also as a key internal component of our algorithm, in order to speed-up the incremental computation of results.   
 

In this as-you-type search setting, it is necessary to serve in a short (fixed) lapse of time, with each keystroke and in social-aware manner,  top-$k$ results  matching the query in its current form, i.e., the terms $t_1, \dots ,t_{r-1}$, and all possible completions of the term $t_r$.  This must be ensured independently of the execution configuration, data features, or scale.  This is why we ensure that our algorithms have also an \emph{anytime} behaviour, being able to  output the  most likely result based on all the preliminary information obtained until a given time limit for the  \algoname\ run is reached.


Our algorithmic solution is validated by extensive experiments for effectiveness, feasibility, and scalability.  Based on data from the Twitter and Tumblr micro-blogging platforms,  two of the most popular  social applications today, we illustrate the usefulness of our techniques for keyword search for microblogs. Based on reviews from Yelp,  we also experiment with keyword search for businesses.

The paper is organised as follows. In Section~\ref{sec:related} we discuss the main related works. We lay out our data and query model in Section~\ref{sec:model}. Our technical contribution is described in Section~\ref{sec:algo} and is evaluated experimentally in Section~\ref{sec:experiments}. We conclude and discuss follow-up research in Section~\ref{sec:conclusion}. For space reasons, more experiments and discussions can be found in a technical report \cite{Lag15}.

\section{Related Work}
\label{sec:related}

Top-$k$ retrieval algorithms, such as the Threshold Algorithm (TA) and the No Random Access algorithm (NRA)~\cite{Fagin01}, which are \emph{early-termination},  have been adapted to network-aware query models for social applications, following the idea of biasing results by the social links, first in~\cite{Amer-Yahia08,Schenkel08}, and then in~\cite{Maniu13-1} (for more details on personalized search in social media we refer the interested readers to the references within~\cite{Maniu13-1,Schenkel08}).

As-you-type (or typeahead) search and query auto-completion are two of the most important features in search engines today, and could be seen as facets of the same paradigm: providing accurate feedback  to queries on-the-fly, i.e.,  as they are being typed (possibly with each keystroke).  In as-you-type search,  feedback  comes in the form of the most relevant answers for the query typed so far, 
allowing some terms (usually, the last one in the query sequence) to be prefix-matched. In query auto-completion, a list of the most relevant query candidates is to be shown for selection, possibly with results for them. We discuss each of these directions separately.

The problem we study in this paper, namely top-$k$ as-you-type search for multiple keywords, has been considered recently  in~\cite{Li:2012}, in the absence of a social dimension of the data. There, the authors consider various adaptations of the well-known TA/NRA top-$k$  algorithms of~\cite{Fagin01}, even in the presence of  minor typing errors (fuzzy search), based on standard tries. 
  A similar fuzzy interpretation for full-text search was followed in~\cite{JiLLF09}, yet not in a top-$k$ setting. The techniques of~\cite{LiJLWF10} rely on precomputed materialisation of top-$k$ results, for values of $k$ known in advance.  In~\cite{BastMW08,BastW06}, the goal is finding all the query completions  leading to results as well as listing these results, based on inverted list and suffix array adaptations; however, the search requires a full computation and then ranking of the results.  For structured data instead of full text, type-ahead search has been considered  in~\cite{FengL12} (XML) and in~\cite{LiFL13} (relational data).

Query auto-completion is the second main direction for instant response to queries in the typing, by which  some  top query completions are presented to the user (see for example~\cite{ShokouhiR12,Shokouhi13,Cai2014} and the references therein). This is done either by following a predictive approach, or by pre-computing completion candidates and storing them in trie structures. Probably the best known example today is the one of Google's instant search, which provides both query predictions (in the search box) and results for the top prediction.  Query suggestion goes one step further by proposing alternative queries, which are not necessarily completions of the input one (see for instance~\cite{Vahabi2013, JiangLVN14}).  In comparison, our work does not focus on queries as first-class citizens, but on  instant results to incomplete queries.

Person (or people) search represents another facet of ``social search'', related to this paper, as the task of finding highly relevant persons for a given seeker and keywords.  Usually, the approach used in this type of application is to identify the most relevant users, and then to  filter them by the query keywords~\cite{Potamias09, Bahmani12}. In this area, \cite{Curtiss13} describes the main aspects of the Unicorn system for search over the Facebook graph, including a typeahead feature for user search.  A similar search problem,  finding a sub-graph of the social network that connects two or more persons, is considered under the instant search paradigm in~\cite{WuTG12}.

Several space-efficient trie data structures for ranked (top-$k$) completion have been studied recently in~\cite{Hsu:2013}, offering various space-time tradeoffs,  and we rely in this paper on one of them, namely the completion trie. In the same spirit, data structures for the more general problem of substring matching for top-$k$ retrieval have been considered in~\cite{HonSV09}.

\section{Model}
\label{sec:model}

We adopt in this paper a well-known generic model of social relevance for information, previously considered among others in~\cite{ManiuCautis12, Maniu13-1,Amer-Yahia08,Schenkel08}. In short, the social bias in scores reflects the social proximity of the producers of content with respect to the seeker (the user issuing a search query), where proximity is obtained by some aggregation of shortest  paths (in the social space) from the seeker towards relevant pieces of information.

We consider a social setting,  in which we have a set of items (could be text documents, blog posts, tweets, URLs, photos, etc) ${\cal I}=\{i_{1},\dots,i_{m}\}$, each tagged with one or more  distinct tags from a tagging vocabulary 
${\cal T}=\{t_{1},t_{2},\dots,t_{l}\}$,  by  users from ${\cal U}=\{u_1, \dots, u_n\}$. We denote our set of unique triples by $Tagged(v,i,t)$, each such triple saying that a user $v$ tagged the item $i$ with tag $t$.  $Tagged$ encodes many-to-many relationships: in particular, any given item can be tagged by multiple users , and any given user can tag multiple items.  We also assume that a user will tag a given item with a given tag at most once.

We assume that users form a social network, modeled for our purposes as an undirected weighted graph $G=({\cal U}, E, \sigma)$, where nodes are users and the $\sigma$  function  associates to each edge $e=(u_1, u_2)$ a value in $(0,1]$, called  \emph{the proximity} (social) score  between $u_1$ and $u_2$.  Proximity may come either from explicit social signals (e.g., friendship links, follower/followee links), or from implicit social signals (e.g., tagging similarity), or from combinations thereof.  (Alternatively,  our core social data can be seen as a  tripartite tagging graph, superposed with an existing friendship network.)

In this setting, the classic  keyword search problem can be formulated as follows: given a seeker user $s$, a keyword query $Q=\{t_1, \dots ,t_r\}$ (a set of $r$ distinct terms/keywords) and a result size $k$, the top-$k$ keyword search problem is to compute the (possibly ranked) list of the $k$ items  having the highest scores with respect to $s$ and the query $Q$.  We rely on the following model ingredients to identify query results.
 
 We model by $score(i~|~ s, t)$,  for a seeker $s$, an item $i$, and one tag $t$,  the relevance of  that item for the given seeker  and query term $t$. 
Generally, we assume
\vspace{-1mm}
\begin{equation}
score(i~|~s,t)=h(fr(i~|~s,t)),  \label{equ:ScorePerItem}
\vspace{-1mm}
\end{equation}
where $fr(i~|~ s, t)$ is the \emph{frequency}  of item $i$ for seeker $s$ and tag $t$, and $h$ is a positive monotone function (e.g., could be based on inverse term frequency, BM25, etc). 

Given a query $Q=(t_1, \dots, t_r)$, the overall score of $i$ for seeker $s$ and $Q$ is simply obtained by summing  the per-tag  scores:
\begin{equation}
score(i~|~ s, Q)= \sum_{t_j\in Q}score(i~|~s,t_j).
\label{equ:ScorestSum}
\vspace{-2mm}
\end{equation}
(Note that this reflects an OR semantics, where items that do not necessarily match all the query tags may still be selected.)

\vspace{-2mm}
\paragraph*{Social relevance model} 
In an exclusively social interpretation, we can explicitate the $fr(i~|~ s, t)$ measure by the  \emph{social frequency}  for  seeker $s$,  item $i$, and one tag $t$,  denoted  $sf(i~|~ s, t)$. 
This measure adapts the classic term frequency (tf) measure to account for the seeker and its social proximity to relevant taggers. We consider that each tagger brings her own weight (proximity) to an item's score, and we  define social frequency as follows:
\begin{equation}
  sf(i~|~ s, t)=\sum_{v\in \lbrace v ~|~Tagged(v,i,t))\rbrace} \sigma(s,v).
  \label{equ:ScoresWeightSum}
  \vspace{-1mm}
\end{equation}
Note that, under the frequency definition of  Eq.~(\ref{equ:ScorePerItem}),  we would follow a ranking approach by which information that may match the query terms but does not score on the social  dimension (i.e., is disconnected from the seeker) is deemed entirely  irrelevant.  

\vspace{-2mm}
\paragraph*{Network-aware relevance model}  A more generic relevance model, which does not solely depend on social proximity but is  network-aware, is one that takes into account textual relevance scores as well. For this, we denote by $tf(t, i)$  the \emph{term frequency} of $t$ in $i$, i.e., the number of times $i$ was tagged with $t$, and 
$IL(t)$ is the  inverted list of items for term $t$, ordered by term frequency.

The  frequency score  $fr(i~|~ s, t)$ is defined as a linear combination of the previously described social relevance and the textual score, with $\alpha \in [0,1]$, as follows:
\vspace{-1mm}
\begin{equation}
	fr(i~|~ s, t)=\alpha \times  tf(t, i) +( 1-\alpha) \times sf(i~|~ s, t).
	\label{equ:genericfreq}
	\vspace{-1mm}
\end{equation}
(This formula thus combines the global popularity of the  item with the one among people close to the seeker.) 

\vspace{-2mm}
\paragraph*{Remark} We believe that this simple model of triples for social data is  the right abstraction for quite diverse types of social media.  Consider Tumblr~\cite{DBLP:journals/sigkdd/ChangTIL14}:  one broadcasts posts to followers  and   rebroadcasts incoming posts;  when
doing so, the re-post is often tagged with chosen tags or short descriptions (hashtags). We can thus see a post and all its re-posted
instances as representing one informational item, which may be tagged with various tags by the users broadcasting it.  Text appearing in a blog post can also be interpreted as tags, provided either by the original author or by those who modified it during subsequent re-posts;  it can also be exploited to uncover implicit tags, based on the 
co-occurrence of tags and keywords in text. Furthermore, a post  that is clicked-on in response to a Tumblr search query can be seen as  being effectively tagged (relevant)  for that query's terms.  All this data has obviously a social nature: e.g., besides  existing follower/followee links, one can even use similarity-based links as social proximity indicators. 

\begin{figure}
\includegraphics[scale=0.58]{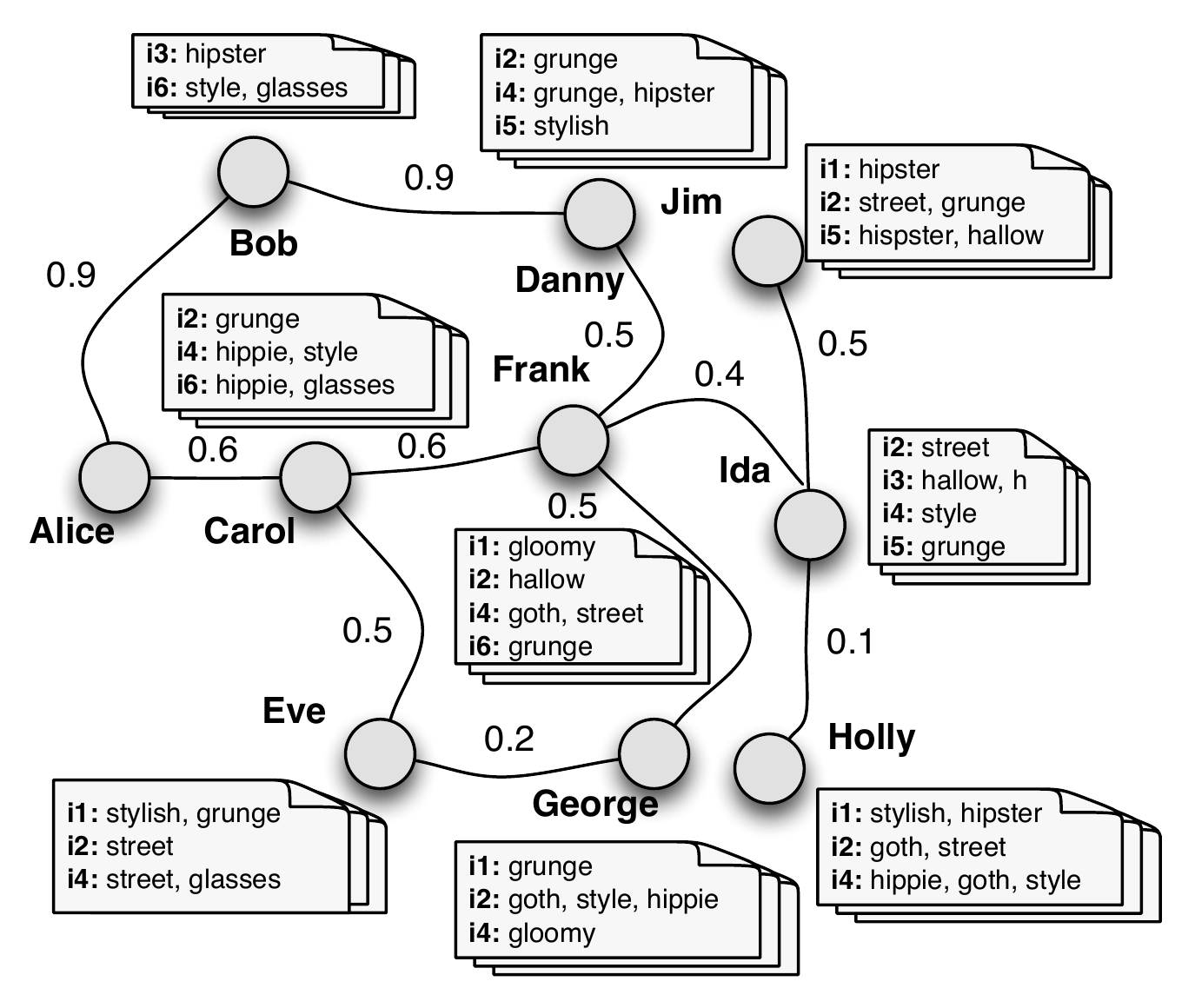}
\vspace{-4mm}
\caption{\small \label{fig:network} Running example: social proximity  \&\ tagging.}
\vspace{-6mm}
\end{figure}

\vspace{-1mm}
\begin{example}
\label{ex:1}
We depict in Figure~\ref{fig:network} a social network and the tagging activity of its users, for a running example based on popular tags from the fashion domain in Tumblr. There, for seeker Alice, we have for instance, for $\alpha=0.2$,  $tf(glasses, i6)  =  2$,
\vspace{-1mm}
\begin{eqnarray*}
sf(i6~|~ Alice, glasses) & =&  \sigma(Alice, Bob) + \sigma(Alice, Carol) \\
& = & 0.9  + 0.6 = 1.5\\
fr(i6~|~ Alice, glasses)  & = & 0.8 \times 1.5 + 0.2 \times 2 
\end{eqnarray*}
\end{example}
\vspace{-2mm}
    

\eat{
\subsection{Threshold algorithms} 
We revisit here the   network-aware retrieval approach of ~\cite{Maniu13-1, Schenkel08}, which belongs to the family of early termination top-$k$ algorithms known as \emph{threshold algorithms}, of which~\cite{Fagin01}'s TA (the Threshold Algorithm) and NRA (No Random-access Algorithm) are well-known examples.   

Generally,  algorithms from this class find the top-$k$ items for an input query $Q$  by  scanning  relevant per-term lists that are ordered descending by relevance,  with inverted lists as a notable example.  During a run, they maintain a set $D$ of already encountered candidate items  $i$, bookkeeping for each candidate a score or  sound upper and lower-bounds on it. 
At each iteration, or at certain intervals,  threshold algorithms may refine these bounds and compare the worst possible score of the $k$th item in $D$  -- assuming the items being ordered in $D$  by their score lower-bounds -- with the \emph{threshold} representing the best possible score of either (i) items in  $D$  outside the top $k$, or (ii) not yet encountered items.  When the threshold is not greater than the worst possible score of $k$th item in $D$, the query execution can terminate, outputting the final top-$k$.  

A major difference between the various threshold algorithms, and in particular between TA and NRA, resides in the way they can access the inverted  lists. The former can do random accesses, as soon as an item has been encountered
while sequentially accessing one of the query terms lists. These random accesses can refine the score of that item from a range to its exact value. On the other hand,  NRA does only  sequential (sorted) accesses, and thus items in the final top-$k$ may be described only by a sound score range instead. 

 In the social-aware retrieval setting, when  social proximity determines the relevance scores, the data exploration must jointly consider the user network (starting from the seeker and visiting users in descending proximity order), the per-user/personal tagging spaces, and all available socially-agnostic index structures such as the  inverted lists. 
  It is therefore important for efficiency to  explore the social network by order of relevance/proximity to the seeker, as to access all the necessary index structures, in a \emph{sequential manner}  as much as possible. We favor such an approach here, instead of an incomplete  ``one dimension at a time'' one,  which  would first rely on one dimension to identify a set of candidate items, and then use the scores  for the other dimension to re-rank or even  filter out some of the candidates.
\inote{why do we favor this approach instead of considering them separately ? Are we going to do experiments on the alternative solution ? or we can cite some paper to say is better. }
        }

\vspace{-3mm}
\paragraph*{Extended proximity} The model described so far takes into account only the immediate neighbourhood of the seeker (the users it connects to explicitly). In order to broaden the scope of the query and go beyond  one's  vicinity in the social network, we  also account for users that are indirectly connected to the seeker, following a natural interpretation that user links  and the query relevance they induce are (at least to some extent) transitive.    To this end, we denote by $\sigma^+$ the resulting measure of  \emph{extended proximity}, which is to be computed from $\sigma$  for any pair of users connected by at least one path in the network.  Now,  $\sigma^+$ can replace $\sigma$ in the definition of social frequency  Eq.~(\ref{equ:ScoresWeightSum}). 


For example, one natural way of obtaining extended proximity scores is by (i) multiplying the weights on a given path between the two users, and  (ii) choosing the maximum value over all the possible paths. Another possible definition for $\sigma^+$  can
rely on an aggregation that penalizes long paths, in a controllable way,  via an  \emph{exponential decay} factor, in the style of the Katz measures for social proximity~\cite{Katz}.  More generally,  any  aggregation function that is monotonically non-increasing over a path, can be used here.  Under this monotonicity assumption, one can browse the network of users \emph{on-the-fly} (at query time) and ``sequentially'', i.e.,  visiting them in the order of their proximity with the seeker.

Hereafter,  when we talk about proximity,  we refer to the extended one, and,  for a given seeker $s$,  the  \emph{proximity vector} of $s$ is the list of users with non-zero proximity with respect to it, ordered decreasingly by proximity values (we stress that this vector is not necessarily known in advance).  

\begin{example}
\label{ex:2}
For example, for seeker Alice, when extended proximity between two users  is defined as the maximal product of scores over paths linking them, the users ranked by proximity w.r.t. Alice are  in order $Bob: 0.9, Danny: 0.81, Carol: 0.6, Frank: 0.4, Eve: 0.3, George: 0.2, Ida: 0.16,   Jim: 0.07, Holly: 0.01$.
\end{example}

\vspace{-6mm}
\paragraph*{The as-you-type search problem} 
We consider in this paper a more useful level of search service for practical purposes, in which queries are being answered \emph{as they are typed}. Instead of assuming that the query terms are given all at once, a  more realistic assumption is  that input queries are sequences of terms $Q=[t_1, \dots ,t_r]$,   in which all terms but the last are to be matched exactly, whereas the last term $t_r$ is to be interpreted as a tag potentially still in the writing, hence matched as a tag prefix. 

We extend the query model in order to deal with tag prefixes $p$ by defining an item's score for $p$ as the maximal one over all possible completions of $p$: 
\begin{eqnarray}
  sf(i~|~ s, p) & = & \max_{t \in \{p's~completions\}} sf(i~|~ s, t) \label{equ:ScorePrefixSF}\\
  tf(p,i) & = & \max_{t \in \{p's~completions\}} tf(t,i)
  \label{equ:ScorePrefixTF}
\end{eqnarray}
(Note that when we compute the importance of an item, we might consider two different tag completions,  for the social contribution and for  the popularity one.) 
\vspace{-1.5mm}
\begin{example}
\label{ex:3}
If Alice's query is \texttt{hipster g}, as \texttt{g}  matches the tags \texttt{gloomy},  \texttt{glasses}, \texttt{goth} and \texttt{grunge}, we have   
\begin{eqnarray*}
  sf(i4~|~ Alice, \texttt{g}) & = & \max_{t \in \{\texttt{g}~~completions\}} sf(i4~|~ Alice, t) \\
  &=  & \max [ sf(i4~|~ Alice, \texttt{gloomy}), \\
  & &  ~~~~~~~~~sf(i4~|~ Alice, \texttt{glasses}),\\
    & &  ~~~~~~~~~sf(i4~|~ Alice, \texttt{grunge}),\\
      & &  ~~~~~~~~~sf(i4~|~ Alice, \texttt{goth})]\\
  &=  & \max [0.2,0.3, 0.81, 0.41] = 0.81
\end{eqnarray*}
\end{example}
\section{\hspace{-1mm}As-you-type search algorithms}
\label{sec:algo}

\begin{figure}[t]
\includegraphics[scale=0.295]{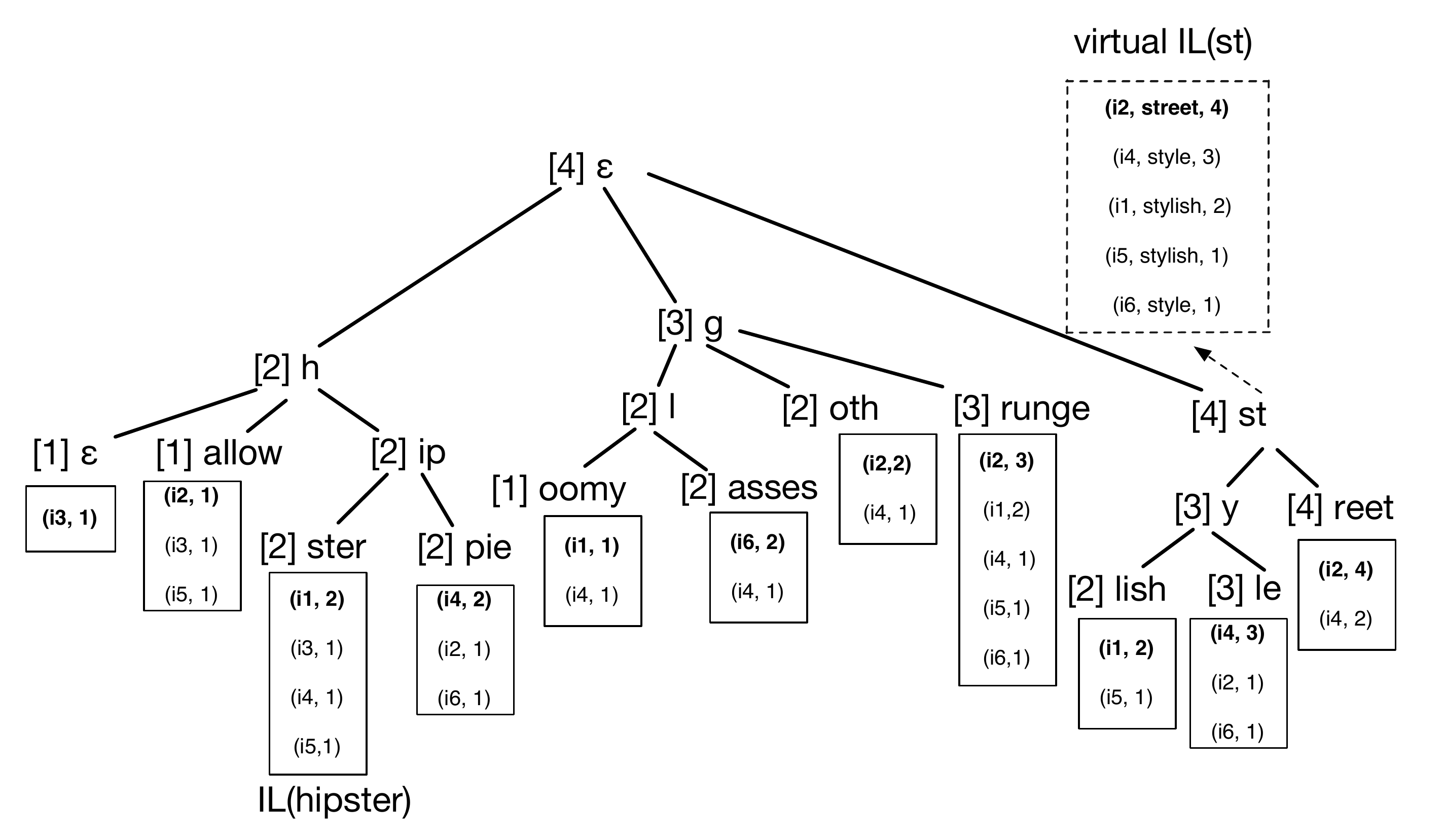}
\vspace{-8mm}
\caption{\small \label{fig:trie} The \indexname\ index.}
\vspace{-5mm}
\end{figure}

We revisit here the  network-aware retrieval approach of ~\cite{Maniu13-1, Schenkel08}, which belongs to the family of early termination top-$k$ algorithms known as \emph{threshold algorithms}, of which~\cite{Fagin01}'s TA (the Threshold Algorithm) and NRA (No Random-access Algorithm) are well-known examples.   



 In the social-aware retrieval setting, when  social proximity determines relevance, the data exploration must jointly consider the network (starting from the seeker and visiting users in descending proximity order), the per-user/personal tagging spaces, and all available socially-agnostic index structures such as inverted lists. 
  It is thus important for efficiency to  explore the social network by order of relevance/proximity to the seeker, as to access all the necessary index structures, in a \emph{sequential manner}  as much as possible. We favor such an approach here, instead of an \emph{incomplete}  ``one dimension at a time'' one,  which  would first rely on one dimension to identify a set of candidate items, and then use the scores  for the other dimension to re-rank or   filter out some of the candidates.

\vspace{-1mm}
\subsection{Non-incremental algorithm}
We first describe the \algoname\ approach for exclusively social relevance  ($\alpha=0$) and  without incremental computation, namely when the full sequence of  terms is given in one keystroke, with the last term possibly a prefix, as $Q=[t_1, \dots ,t_r]$. 
We follow an early-termination approach that is ``user-at-a-time'': its  main loop step visits a  new user and   the items that were tagged by her with  query terms. 
 Algorithm~\ref{alg:soctopktrust} gives the flow of \algoname.

\vspace{-3mm}
\paragraph*{Main inputs} For each user $u$ and tag $t$, we assume a precomputed selection over the \emph{Tagged} relation, giving the items tagged by $u$ with $t$;  we call these the \emph{personal spaces} (in short, p-spaces). No particular order is assumed for the items appearing in a user list.  

We also assume that, for each tag $t$, we have an inverted list $IL(t)$ giving the items $i$ tagged by it, along with their term frequencies $tf(t, i)$\footnote{Even when $\alpha=0$,  although  social frequency does not depend directly on  $tf$ scores, we will exploit the inverted lists and the $tf$ scores by which they are ordered, to better estimate score bounds.}, ordered descending by them. The lists can be seen as unpersonalized indexes. 
 A completion trie over the set of tags represents the access layer to these lists. As in Patricia tries, a node can represent more than one character, and the scores corresponding to the heads of the lists are used for ranked completion: each leaf has the score of the current entry in the corresponding inverted list, and each internal node has the maximal score over its children (see example below). This index structure is denoted hereafter the \indexname\ index.

\begin{figure}[t]
\includegraphics[scale=0.28]{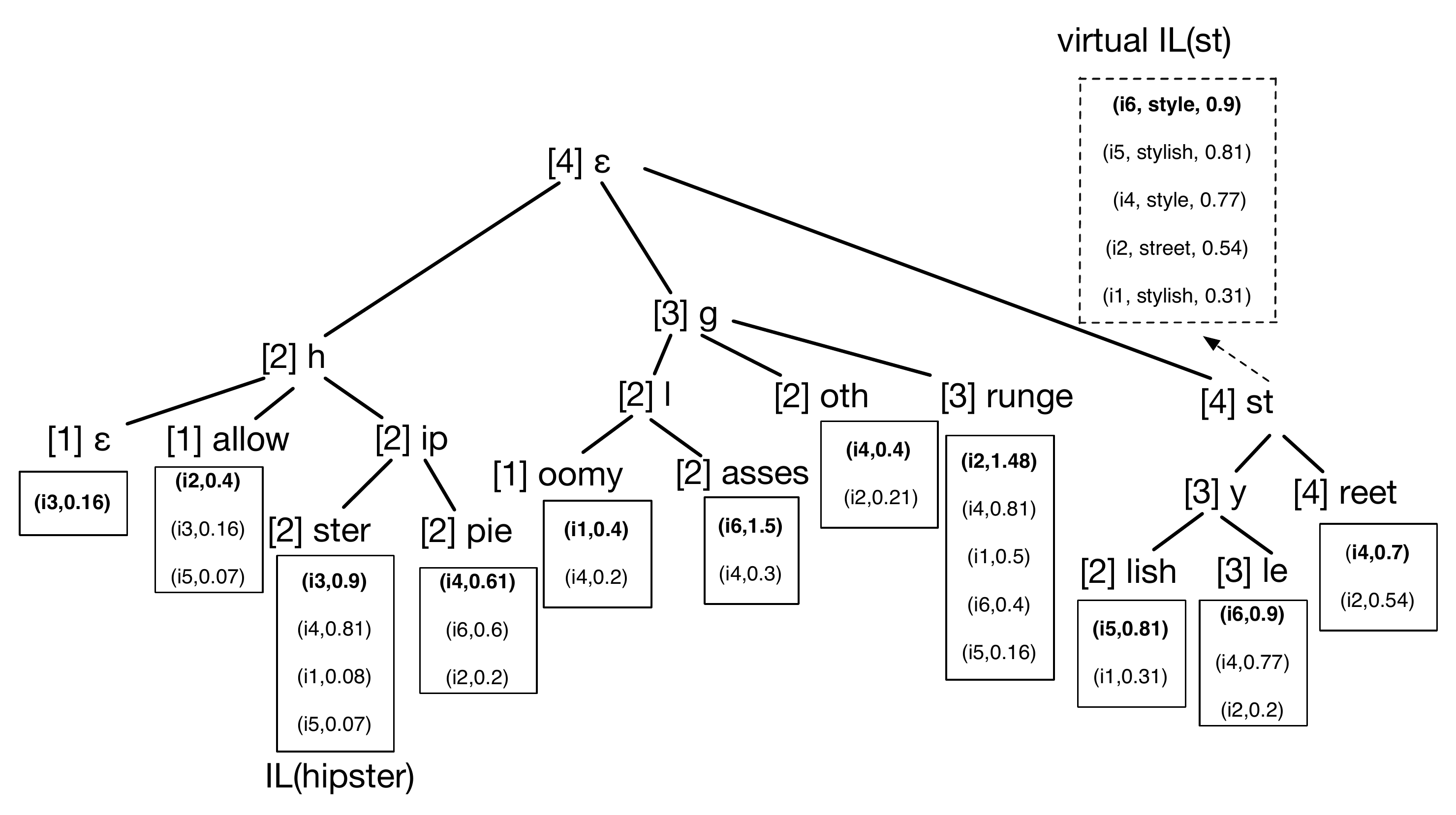}
\vspace{-7mm}
\caption{\small \label{fig:trieALICE} Alice's personalized \indexname\ index.}
\vspace{-5mm}
\end{figure}

\vspace{-1.8mm}
\begin{example}[CT-IL index]
\label{ex:4}
We give in Figure~\ref{fig:trie} an illustration of the main components of \indexname, for our running example. Each of the tags has below it the inverted list (the one of the \texttt{hippie} tag is explicitly indicated).  The cursor positions in the lists are in bold.  By storing the maximal score at each node (in brackets in Figure~\ref{fig:trie}), the best (scoring) completions of a given prefix can be found by using a priority queue, which is initialized with the highest node matching that prefix. With each pop operation, either we get a completion of the prefix, or we advance towards one, and we insert in the queue the children of the popped node.

For comparison, we also illustrate in Figure~\ref{fig:trieALICE} the \indexname\  index that would allow us to process efficiently Alice's top-$k$ queries, without the need to resort to accesses in social network and p-spaces. Obviously, building such an index for each potential seeker would not be feasible. 
\end{example}
\vspace{-2.2mm}

While leaf nodes in the trie correspond to concrete inverted lists, we can also see each internal node of the trie and the corresponding keyword prefix as described by  a ``virtual inverted list'', i.e., the ranked union of all inverted lists below that node. As defined in  Eq.~(\ref{equ:ScorePrefixTF}),~(\ref{equ:ScorePrefixSF}), for such a union, for an item appearing in entries of several of the unioned lists, we keep only the highest-scoring entry. In particular, for the term $t_r$ of the query, by $IL(t_r)$ we refer to the virtual inverted list corresponding to this tag prefix. There is one notable difference between the concrete inverted lists and the virtual ones: in the former, entries can be seen (and stored) as pairs $(item, score)$ (the tag being implied); in the latter, entries must be the form $(item, tag, score)$, since different tags (completions) may appear in such a  list. 

For each $t \in \{t_1, \dots ,t_{r}\}$, we denote by  $top\_item(t)$  the item present at the current (unconsumed) position of $IL(t)$,  we use $top\_tf(t)$ as short notation for the term frequency associated with this item, and, for $IL(t_r)$, we also denote by $top\_tag(t_r)$ the $t_r$ completion in   the current entry. 

\vspace{-2mm}
\begin{example}[Virtual lists]
\label{ex:5}
The virtual inverted list for the prefix \texttt{st} is given in Fig.~\ref{fig:trie}. The $top\_tag(\texttt{st})$ is \texttt{street}, for $top\_item(\texttt{st})$ being $i2$, for its  entry scored $4$ dominates the one scored only $2$, hence with a $top\_tf(\texttt{st})$ of $4$. A similar one, for the  ``personalized''  \indexname\ index for seeker Alice is given in Fig.~\ref{fig:trieALICE}.
\end{example}
\vspace{-3mm}

\vspace{-3mm}
\paragraph*{Candidate buffers} For each tag $t \in \{t_1, \dots ,t_{r-1}\}$, we keep a list $D_t$ of candidate items $i$, along with a sound score range: a lower-bound and an upper-bound for $sf(i ~|~ s, t)$ (to be explained hereafter). Similarly, in the case of  $t_r$, for each completion $t$  of $t_r$ already encountered during the query execution in p-spaces (i.e., by triples $(u,i, t)$  read in some $u$'s p-space), we  record in a $D_{t}$ list the candidate items and their score ranges.  Candidates in  these  $D$-buffers are  sorted in descending order by their score lower-bounds.

An item becomes candidate and is included in $D$-buffers only when it is first met in a $Tagged$ triple matching a query term.

 For uniformity of treatment,  a special item  $*$ denotes  all the yet unseen items, and it implicitly appears in each of the $D$-lists; note that, in  a given $D_t$ buffer,   $*$ represents both  items which are not yet candidates, but also candidate items which may already be candidates but  appear only in other $D$-buffers (for tags other than $t$).

\vspace{-3mm}
\paragraph*{Main algorithmic components} When accessing the \indexname\ index, inverted list entries are consumed in some $IL(t)$ only when the items they refer to are candidates (they appear in at least one $D_{t'}$ buffer, which may not necessarily  be $D_t$ itself)\footnote{The rationale is that our algorithm does not make any ``wild guesses'', avoiding reads that may prove to be irrelevant and thus leading to sub-optimal performance.}. We keep in lists called $CIL_t$ (for \emph{consumed IL entries}) the items read (hence known candidates)  in the inverted lists (virtual or concrete), for $t$ being either in $\{t_1, \dots ,t_{r-1}\}$ or a completion of $t_r$ for which a triple $(item, t, score)$ was read in the virtual list of $t_r$.    We also record by the set  $C$  all  $t_r$ completions  encountered so far in p-spaces.  We stress that the $t_r$ completions encountered in p-spaces may not necessarily coincide with those encountered in $IL(t_r)$.

 For each $t$ being either in $\{t_1, \dots ,t_{r-1}\}$ or a completion of $t_r$ already in $C$, by $unseen\_users(i,t)$ we denote the maximal number of yet unvisited users who may have tagged item $i$ with tag $t$. This number is initially set to the maximal possible term frequency of $t$ over all items. $unseen\_users(i,t)$ then reflects at any moment during the run of the algorithm  the difference between the number of taggers of $i$ with $t$ already visited and one of either
 \begin{itemize}
 \vspace{-1mm}
  \item the value $tf(t,i)$, if this term frequency has been read already by accessing \indexname,  or otherwise
   \vspace{-1mm}
 \item  the value $top\_tf(t)$, if $t \in \{t_1, \dots ,t_{r-1}\} $, or
  \vspace{-1mm}
 \item the  value $top\_tf(t_r)$,  if $t$ is instead a completion of $t_r$.
  \vspace{-1mm}
 \end{itemize}
 
 During the algorithm's run, for known candidates $i$ of some $D_t$, we accumulate in $sf(i ~|~ s, t)$  the social score (initially set to $0$).

 Each time we visit a user $u$ having a triple $(u, i,t)$ in her p-space (Algorithm~\ref{alg:pspace}), we can 
 \begin{enumerate}
  \vspace{-1mm}
 \item update $sf(i ~|~ s, t)$  by adding $\sigma^+(s,u)$ to it, and 
  \vspace{-1mm}
 \item decrement $unseen\_users(i,t)$;  when this value reaches $0$,  the social frequency value $sf(i ~|~ s, t)$ is final. 
  \vspace{-1mm}
 \end{enumerate}
 
 The maximal proximity score of yet to be visited users is denoted $max\_proximity$.  With this proximity bound, a sound score range for candidates $i$ in $D_t$ buffers is computed and maintained as  
   \begin{itemize}
    \vspace{-1mm}
   \item  a score upper-bound (maximal score) \textsc{MaxScore}$(i~|~s, t)$,   by 
	$max\_proximity \times unseen\_users(i,t) +  sf(i ~|~ s, t)$. 
	 \vspace{-1mm}
\item    a score lower-bound (minimal  score), \textsc{MinScore}$(i~|~s,t)$,  by assuming that the current social frequency $sf(i ~|~ s, t)$ is the final one (put otherwise, all remaining taggers $u$ of $i$ with $t$, which are yet to be encountered, have  $\sigma^+(s,u)=0$).    
   \vspace{-1mm}
   \end{itemize}
  
  The interest of consuming the  inverted list entries (Algorithm~\ref{alg:index}) in \indexname,  whenever top items  become candidates, is to keep  as accurate as possible the worst-case estimation on the number of unseen taggers.  
  Indeed, when such a tuple $(i, t, score)$ is accessed, we can do some adjustments on score estimates: 
  \begin{enumerate}
   \vspace{-1mm}
  \item if $i \in D_t$, we can mark the number of unseen taggers of $i$ with $t$ as no longer an estimate but an \emph{exact} value; from this point on, the number of unseen users will only change whenever new users who tagged $i$ with $t$ are visited,
  \vspace{-1mm}
  \item  by advancing  to the next best item in  $IL(t)$, for $t \in \{t_1, \dots ,t_{r-1}\} $, we can refine  the $unseen\_users(i',t)$  estimates for all candidate items $i'$ for which the exact number of users who tagged them with $t$ is yet unknown,  
 \vspace{-1mm}
 \item by advancing  to the next best item in  $IL(t_r)$,  with some $t = top\_tag(t_r)$  completion of $t_r$,  if $t \in C$, we can refine the estimates  $unseen\_users(i',t)$ for all candidate items $i' \in D_t$ for which the exact number of users who tagged them  with $t$ is yet unknown.  
 \vspace{-1mm}
  \end{enumerate}

     \vspace{-5mm}
  \paragraph*{Termination condition} From the per-tag $D_t$ buffers, we can infer lower-bounds on the global score w.r.t. $Q$ for a candidate item (as defined in Eq.~(\ref{equ:ScorestSum}))  by summing up its score lower-bounds from $D_{t_1}, \dots, D_{t_{r-1}}$ and its maximal score lower-bound across all $D_t$ lists, for completions $t$ of $t_r$. Similarly,  we can infer an upper-bound on the global score  w.r.t. $Q$ by summing up  score upper-bounds from $D_{t_1}, \dots, D_{t_{r-1}}$ and the maximal upper-bound across all $D_t$ lists,  for completions $t$.  
  
  After sorting the candidate items (the wildcard item included) by their global score lower-bounds,  \algoname\  can terminate whenever (i) the wildcard item is not among the top-$k$ ones, and  (ii) the score upper-bounds of items not among the  top-$k$  ones are less than the score lower-bound of the $k$th item in this ordering (we know that the top-$k$ can no longer change). 

    \setlength{\textfloatsep}{0pt}
    
 \begin{algorithm}[!t]
\caption{$\algoname$ (non-incremental, for $\alpha=0$)}
	\begin{algorithmic}[1]\small
	\REQUIRE\emph{ seeker $s$, query $Q=(t_1, \dots, t_r)$}
	\FORALL{\emph{users $u$}}
		\STATE $\sigma^+(s,u)=-\infty$
	\ENDFOR
	\FORALL{ tags $t \in \{t_1, \dots ,t_{r-1}\}$ }
		\STATE $sf(i ~|~ s, t)= 0$
		\STATE $D_t=\emptyset$, $CIL_t=\emptyset$
		\STATE \emph{set $IL(t)$ position on first entry}
	\ENDFOR
	\STATE \emph{set $IL(t_r)$ position on first entry}
	\STATE $\sigma^+(s,s)=0$; 
	\STATE $C = \emptyset$ ($t_r$ completions)
	\STATE $H\gets$ \emph{priority queue on users; init.  $\{s\}$, computed on-the-fly}
	\WHILE{$H\neq \emptyset$}
	\STATE u=\textsc{extract\_max}(H);
		\STATE \textsc{process\_p\_space}(u);
			\STATE \textsc{process\_\indexname};
		\IF{termination condition}
			\STATE \textbf{break}
		\ENDIF
	\ENDWHILE
	\RETURN top-$k$ items 
	\end{algorithmic}
\label{alg:soctopktrust}
\end{algorithm}    

 \begin{algorithm}[t]
\caption{ \textsc{Subroutine process\_p\_space}(u) }
	\begin{algorithmic}[1]\small
\FORALL{\emph{tags $t \in \{t_1, \dots ,t_{r-1}\}$, triples $Tagged(u,i,t)$}}
					\IF{$i \not \in D_t$} 
			\STATE \emph{add $i$ to $D_t$} 
			\STATE $sf(i ~|~ s, t)\gets 0 $
			\STATE $unseen\_users(i, t) \gets top\_tf(t)$ 
		 				\ENDIF		
					\STATE $unseen\_users(i, t) \gets unseen\_users(i,t) - 1$		
					\STATE $sf(i ~|~ s, t)\gets sf(i ~|~ s, t) + \sigma^+(s,u)$	
		\ENDFOR
		\FORALL{\emph{tags $t$ completions of $t_r$, triples $Tagged(u,i,t)$}}
		\IF{$t \not \in C$}
		\STATE \emph{add $t$ to $C$, $D_t=\emptyset$ }
		\ENDIF 
			\IF{$i \not \in D_t$} 
			\STATE \emph{add $i$ to $D_t$} 
			\STATE $sf(i ~|~ s, t)\gets 0 $
			\STATE $unseen\_users(i, t) \gets top\_tf(t)$ 
	 				\ENDIF		
					\STATE $unseen\_users(i, t) \gets unseen\_users(i,t) - 1$		
					\STATE $sf(i ~|~ s, t)\gets sf(i ~|~ s, t) + \sigma^+(s,u)$	
		\ENDFOR
\end{algorithmic}
\label{alg:pspace}
\end{algorithm}    

As  in~\cite{Maniu13-1}, it can be shown that $\algoname$ visits users who may be relevant for the query in decreasing proximity order and, importantly, that it visits  as few users as possible (it is  \emph{instance optimal} for this aspect, in the case of exclusively social relevance).

\begin{algorithm}[t]
\caption{ \textsc{Subroutine process\_\indexname}}
	\begin{algorithmic}[1]\small
  \WHILE{\emph{$\exists t \in  Q$ s.t. $i=top\_item(t) \in \bigcup_x D_x$}}
		  \IF{$t\neq t_r$}
	   \STATE {$tf(t,i) \gets top\_tf(t)$}    \hspace{-0.7mm} ($t$'s frequency in $i$ is now known)
	   \STATE {\emph{advance $IL(t)$ one position}}
		\STATE {$\Delta \gets  tf(t,i) -  top\_tf(t)$ (the top\_tf drop)}
		\STATE{\emph{add $i$ to $CIL_t$}}	
	\FORALL{\emph{items $i'\in D_t \setminus CIL_t$}}
	\STATE $unseen\_users(i', t) \gets unseen\_users(i',t) - \Delta$
	\ENDFOR		
	\ENDIF
	  \IF{$t = t_r$}
	   \STATE {$t' \gets top\_tag(t_r)$}   (some $t_r$ completion $t'$)
	  \STATE {$tf(t',i) \gets top\_tf(t_r)$}    \hspace{-0.9mm} ($t'$'s frequency in $i$  known)
	   \STATE {\emph{advance $IL(t_r)$ one position}}
		\STATE {$\Delta \gets  tf(t',i) -  top\_tf(t_r)$ (the top\_tf drop)}
		\STATE{\emph{add $i$ to $CIL_{t'}$ or set $CIL_{t'}$ to \{i\} if previously empty}}	
	\FORALL{\emph{$t'' \in C$ and items $i'\in D_{t''} \setminus CIL_{t''}$}}
	\STATE $unseen\_users(i', t'') \gets unseen\_users(i',t'') - \Delta$
	\ENDFOR		
	\ENDIF
		   \ENDWHILE
\end{algorithmic}
\label{alg:index}
\end{algorithm}    

\vspace{-2mm}
 \begin{example}
Revisiting our running example, let us assume Alice requires the top-$2$ items for the query $Q=[\texttt{style}, \texttt{gl}]$ ($\alpha=0$). The first data access steps of \algoname\ are as follows: at the first execution of the main loop step, we visit $Bob$, get his p-space, adding $i6$ both to the $D_{\texttt{style}}$ buffer and to a $D_{\texttt{glasses}}$ one. There may be at most two other taggers of $i6$ with \texttt{style} ($unseen\_users(i6,\texttt{style})$), and at most one other tagger of $i6$  with  \texttt{glasses} ($unseen\_users(i6,\texttt{glasses})$). No reading is done in $IL(\texttt{style})$, as its current entry gives the non-candidate item $i4$,  but we can advance with one pop in the virtual list of the $\texttt{gl}$ prefix, for candidate item $i6$. This clarifies that there is  \emph{exactly} one other tagger with  \texttt{glasses} for $i6$. After this read in the virtual list of $\texttt{gl}$, we have $top\_item(\texttt{gl}) = i1$ (if we assume  that items are also ordered by their ids). 
 At this point $max\_proximity$ is $0.81$. Therefore, we have   
 \begin{eqnarray*}
 \textsc{MaxScore}(i6~|~Alice, \texttt{style}) & = &	0.81 \times 2 +  0.9 \\
 \textsc{MinScore}(i6~|~Alice, \texttt{style}) & = &	0.9 \\
 \textsc{MaxScore}(i6~|~Alice, \texttt{glasses}) & = &	0.81 \times 1 +  0.9  \\
  \textsc{MinScore}(i6~|~Alice, \texttt{glasses}) & = &	0.9 
   \vspace{-1mm}
 \end{eqnarray*}
We thus have that $score(i6|Alice, Q)$ is between $1.8$ and $4.23$. 


At the second execution of the main loop step, we visit $Danny$, whose p-space does not contain relevant items for $Q$. A  side-effect of this step is that  $max\_proximity$ becomes $0.6$, affecting the upper-bound scores above: $score(i6~|~Alice, Q)$ can  now be estimated between $1.8$ and $3.6$.

At the third execution of the main loop step, we visit $Carol$, and find the relevant p-space entries for $i4$ (with tag \texttt{style}) and $i6$ (with tag \texttt{glasses}). Now  $max\_proximity$ becomes $0.4$. Also,  we can  advance with one pop in the inverted list of $\texttt{style}$. This clarifies that there are \emph{exactly} 2 other taggers with  \texttt{style} on $i4$,  and now we have $top\_item(\texttt{gl}) = i1$ and  $top\_item(\texttt{style}) = 2 $. This makes  $score(i6~|~Alice, Q)$ to be known precisely at $2.4$,   $score(i4~|~Alice, Q)$ to be estimated between $0.6$ and $0.6 + 3 \times 0.4=1.8$, and $score(*~|~Alice, Q)$ is at most $0.8$. 
  
  \label{ex:6}
  \end{example}

\subsection{Adaptations for the network-aware case}

Due to lack of space, we only sketch in this section the necessary extensions  to Algorithm~\ref{alg:soctopktrust} for arbitrary $\alpha$ values, hence for any textual-social relevance balance.  When $\alpha \in [0,1]$, at each iteration, the algorithm can alternate  between two possible execution branches: the \emph{social branch} (the one detailed in Algorithm~\ref{alg:soctopktrust}) and a  \emph{textual branch}, which is a direct adaptation of  NRA  over the \indexname\  structure, reading in parallel in all the query term lists (concrete or virtual). Now, items can become candidates even without being encountered in p-spaces, when read in inverted lists  during an execution of the textual branch. As before, each read from \indexname\ is associated with updates on score estimates such as $unseen\_users$. For a given item $i$ and tag $t$, the maximal possible $fr$-score can be obtained by adding to the previously seen  maximal possible $sf$-score (weighted now by $1-\alpha$) the maximal possible value of $tf(t,i)$; the latter may be known (if read in \indexname), or estimated as  $top\_tf(t)$ otherwise.  Symmetrically,  the minimal possible value for $tf(t,i)$ is used for lower bounds; if not known, this can be  estimated as the number of visited users who tagged $i$ with $t$. 

The choice between the two possible execution branches can rely on heuristics which estimate their utility w.r.t approaching  the final result. Two such heuristics are explained in~\cite{Maniu13-1,Schenkel08},  guiding this choice either  by estimating the maximum potential score of each branch, or by  choosing the branch that is the most likely to refine the score  of the item outside the current top-$k$ which has the highest estimated score (a choice that  is likely to advance the run of the algorithm closer to termination). 
\vspace{-1.5mm}
\subsection{Adaptations for  incremental computation}
\vspace{-0.5mm}
We extended the approach described so far to perform the as-you-type computation \emph{incrementally}, as follows: 
\begin{enumerate}
\vspace{-2mm}
\item when a new keyword is initiated (i.e., $t_r$ is one character long), we take the following steps in order: 
\vspace{-1mm}
\begin{enumerate}
\vspace{-1mm}
\item purge all $D_t$ buffers  for $t \in C$, except for $D_{t_{r-1}}$ ($t_{r-1}$ is no longer a potential prefix, but a complete term),
\vspace{-1mm}
\item reinitialize $C$ to the empty set,
\vspace{-1mm}
\item purge all $CIL_t$ buffers  for $t \not \in \{t_1, \dots,  t_{r-1}\}$,
\vspace{-1mm}
\item reinitialize the network exploration (the queue $H$) to start from the seeker, in order to visit again p-spaces looking for triples for the new prefix, $t_r$.
(This amounts to the following  changes in Algorithm~\ref{alg:soctopktrust}: among its initialisation steps (1-12), the steps (4-8) are removed, and new steps for points (a) and (c) above are added.)
\vspace{-1mm}
\end{enumerate} 
\vspace{-2mm}
\item when the current  $t_r$ is augmented with one additional character (so $t_r$ is at least two characters long), we  take the following steps in order: 
\vspace{-1mm}
\begin{enumerate}
\vspace{-1mm}
\item purge $D_t$ buffers for $t \in C$ s.t. $t$ is not a $t_r$ completion 
\vspace{-1mm}
\item remove from $C$ all $t$s which aren't completions for $t_r$,
\vspace{-1mm}
\item purge all $CIL_t$ buffers  for $t \not \in \{t_1, \dots t_{r-1}\} \cup C$,
\vspace{-1mm}
\item resume the network exploration.

(This amounts to the following  changes in Algorithm~\ref{alg:soctopktrust}: among its initialisation steps (1-12), the steps (4-8) and (10-12) are removed, and new steps for points (a), (b), and (c) above are added.)
\vspace{-1mm}
\end{enumerate} 
\end{enumerate}
Note that, in the latter case, we can efficiently do the filtering operations by relying on a simple trie structure for directly accessing the data structures ($D$-lists, $CIL$-lists, the $C$ subset) that remain valid for the new prefix.  

\eat{
\subsection{Materializing virtual lists in \indexname}
We discuss in this section one important optimisation for  query execution over the \indexname\  index structure.  When some of the virtual lists corresponding to internal trie nodes in \indexname\ are \emph{materialized}, this can speed-up the sorted access time, since heap-based operations are  avoided. We have a clear tradeoff between the necessary memory space for this materialization and the potential gains at query time. (Offline index construction costs can also be taken into account, but we see them as a less important factor for practical scenarios.) Therefore, the lists to be materialized must be chosen carefully, in order to make the most of the available memory budget. 

We describe next our strategy for choosing virtual lists/prefixes for materialization. For presentation purposes, this is discussed by  assuming first that the optimisation is dedicated to a single potential seeker, $s$. 

Given  a prefix $p$ in the completion trie, the main new ingredients are the following:
\begin{itemize}
\item the allotted space budget $B$,
\item a query likelihood: the probability for $s$ to query by prefix $p$, denoted $prob(s,p)$; this can be inferred either by analysing the p-space of $s$, her previous query logs, or directly from the trie when no such information is available,
\item the trie subtree rooted at the node corresponding to $p$, denoted $\indexname(p)$,
\item the number of concrete inverted lists in $\indexname(p)$ (the number of leaves in this subtree), $NbLeaves(p)$; note that the higher this number, the more useful the materialization of $p$ will be.
\item for the fixed seeker $s$,  a sequence over all items tagged with completions of $p$ by taggers reachable from $s$, denoted $ItemVisit(s,p)$. In this sequence, each item $i$ appears at  the rank of tagger of $i$ with $p$ completions who is closest to $s$; put otherwise, this sequence describes the order in which items tagged by $p$ completions would be discovered in a social network exploration by  \algoname. 
\end{itemize}
\begin{example}
For example, $ItemVisit(Alice,st)$ is .... 
\end{example}
The utility of materializing a list $IL(p)$ can be described by the number of accesses to it, during \algoname. For a fixed limit $n$ on the number of visited users, which could for instance be estimated from a query execution time limit such as $100ms$, let $ItemVisit^{1,n}(s,p)$ denote $ItemVisit(s,p)$'s subsequence up to rank $n$. The number of access in $IL(p)$ for seeker $s$ under limit $n$, denoted $NbAccesses(s,n,p)$ is then equal to the number of consecutive  $IL(p)$ entries, starting with the first one,  that belong to $ItemVisit^{1,n}(s,p)$. 

We have implemented a cost-based approach which, for a given seeker $s$ and limit  $n$, over all possible prefixes $p$,  decides greedily which virtual lists should be materialized, by  taking  into account the budget size $B$, the likelihood values $prob(s,p)$, the virtual list sizes $|IL(p)|$, the values of $NbAccesses(s,n,p)$, the sizes of  trie subtrees $\indexname(p)$, and the values of $NbLeaves(p)$.

When all seekers are to be taken into account, in order to choose the lists to be materialized, a direction we intend to evaluate in the future is one by which an aggregated utility score for each prefix $p$ is obtained, either (i) with respect to all seekers, weighted by the likelihood scores for each of them, or (ii) with respect to some top-$l$ seekers, by likelihood.  Based on such scores, the most promising prefixes can then be  chosen greedily.
}

\eat{
\subsection{Finding the most probable top-$k$ anytime} 
As argued before, we also see as crucial  for the as-you-type search approach to have an \emph{anytime} behaviour, in the following sense: it should explore the social space and existing data structures/indexes  in the most efficient manner, maintaining the candidate buffers, until a time limit is met or an external event occurs. Indeed, in practice, we can expect that most searches will not meet the termination condition within the imposed time limit; 
when this happens, it should output the most likely top-$k$ result, as formalised next. 

Working with the intermediate result at any step in the \algoname\ computation, in particular  the $D$-buffers, if an exact top-$k$
cannot yet be outputted, a most informative
result would consist of two disjunctive, possibly-empty sets
of items among the candidates appearing in $D$:  a set of all the items that must  be in the final  top-$k$ result, and a set of all items that may also be in the final top-$k$. More precisely, the former set of \emph{guaranteed} items are simply those for which at most $k-1$ items can be inferred as having a  score upper-bound that  is greater than their score lower-bound. The latter set of \emph{possible} items are simply those that do not verify the previous condition and for which at most $k-1$ items are inferred as having a  score lower-bound that  is greater than their score  upper-bound. 

Furthermore, from the set of possible items, we can complete the set of guaranteed ones with those that are the most likely  to make it to the final top-$k$, as follows: assuming a known probability density
function (e.g, the uniform one) for scores within the known bounds, we  can reason about the likelihood of a top-$k$ selection. In practice, this can be implemented efficiently via a sampling-based approach~\cite{Maniu13-2}. Obviously, this result represents only an approximation for the one that would be obtained if running the retrieval algorithm until the threshold condition is met.   \inote{check}

We can use in \algoname\ a  sub-routine that is an adaptation of the SR-TA procedure (for Score-Ranges Threshold Algorithm) of~\cite{Maniu13-2}. By examining the $D$-buffers of candidate items (called views in~\cite{Maniu13-2}), it identifies (i) the items that are certainly in the top-$k$ result, and then (ii) the items  which are the most likely to complete the top-$k$ result. We stress that our sub-routine for this task is a threshold method as well, since it aims to avoid reading all the information in order to produce its output, following the generic TA flow (for space reasons, further details are omitted here):
\begin{itemize}
\item taking as input all the $D$-buffers, which are assumed to be sorted by score lower-bounds,
\item sequential reads are done in  round-robin manner in the buffers, and
\item each read of an item $i$ in a buffer $D_t$ is complemented by random reads to find $i$'s existing score ranges in all other $D$-buffers, for tags other than $t$.
\end{itemize}
Concretely, this sub-routine could  be called at  the last line in Algorithm~\ref{alg:soctopktrust}, whenever the exact top-$k$ is not available, when the main loop execution reaches the time limit.  

However, one particularity of our setting is that the number of $D$-buffers may be  large (if $C$ is large), which means that each lookup for an item $i$'s score ranges may have an important cost. This is why a different organization, which is ``per-item'' instead of ``per-tag'',  for information in buffers $D_t$ for $t\in C$, is more suited here. For each item $i$, we can keep in a trie structure the $t_r$ completions $t$ for which  triples $(user,i,t)$ have been encountered in p-spaces so far. Each leaf node provides the score range for that item-tag pair. The cost of finding $i$'s maximal lower-bound and maximal upper-bound over all completions by which $i$ has been tagged can be reduced by employing a completion trie as well:  let $D-trie(i)$ denote this index, built for each candidate $i$ tagged by at least one $t_r$ completion. Using $D-trie(i)$, we can in constant time access $i$'s score bounds for the $t_r$ term, and also zoom-in on the part of the trie that remains relevant when $t_r$ receives a new character.

\begin{example}
\label{ex:9}
\inote{we put one here?}
\end{example}

\inote{explain extension of the algorithm for alpha non-zero}
}

\vspace{-1mm}
\subsection{Finding the most likely top-$k$ anytime} 
As argued before,  we also see as crucial  for the as-you-type search approach to have an \emph{anytime} behaviour, in the following sense: it should explore the social space and existing data structures / indexes  in the most efficient manner, maintaining the candidate buffers, until a time limit is met or an external event occurs. Indeed, in practice, we can expect that \emph{most searches will not meet the termination condition within the imposed time limit}; 
when this happens, we must output the most likely top-$k$ result. In our case, this can be easily obtained from the intermediate result at any step in the \algoname\ computation, in particular  the $D$-buffers, e.g., by adapting the more general SR-TA procedure (for Score-Ranges Threshold Algorithm) of~\cite{Maniu13-2}, especially for the fact that we may have many $D$-buffers (if $C$ is large). This calls for  a different organization, which is ``per-item'' instead of ``per-tag'',  for information in buffers $D_t$ for $t\in C$. In short, for each item $i$, we can keep in a trie structure the $t_r$ completions $t$ for which  triples $(user,i,t)$ have been encountered in p-spaces so far, with each leaf providing the score range for that item-tag pair. Further details are omitted here. 


\eat{

\begin{table}[th]
\begin{center}
\begin{tabular}{  l | r | r  | r | r   }
\hline
&  {\bf Twitter}  &  {\bf Amazon} &  {\bf Tumblr} &  {\bf Yelp}\\
\hline
Nb. of unique users 			& \small   $458,117$ 	& \small  $886,814$ &  \small $612,425$ & \small  $29,293$ \\ 
Nb. of unique items 			&  \small $1.6M$  & \small  $252,891$	&  \small $1.4M$ &  \small $18,149$ \\ 
Nb. of unique tags 			&  \small $550,157$ 	& \small  $91,352$ &  \small $2.3M$ &  \small $177,286$  \\ 
Nb. of triples 				&  \small $13.9M$ 	&  \small $24.6M$ & \small  $11.3M$  &  \small $30.3M$ \\ 
Avg. nb. of tags per item  	&  \small $8.4$ 	& \small  $97.5$ & \small  $7.9$ &  \small  $685.7$  \\ 
Avg. tag length				& $$ \\
\hline
\end{tabular}
\end{center}
\vspace{-3mm}
\caption{Statistics on the datasets we used in our experiments.
}
\label{tab:StatDataset}
\end{table}

}

\section{Experiments}
\label{sec:experiments}

We evaluate in this section the effectiveness, scalability and efficiency of the  \algoname\ algorithm.
We used a Java implementation of our algorithms, on a low-end Intel Core i7 Linux machine with 16GB of
RAM. We performed our experiments in an all-in-memory setting, for datasets of medium size (10-30 millions of tagging triples). We  describe first the applications and datasets we used for evaluation. 

\vspace{-1mm}
\subsection{Datasets}
\label{sec:dataset}
\vspace{-1mm}
\noindent We used several popular social media platforms, namely  Twitter, Tumblr, and Yelp,   from which we built corresponding sets of (user, item, tag) triples.  Table \ref{tab:StatDataset} reports some statistics about each dataset.

\begin{table}[th]
\begin{center}
\begin{tabular}{  l | r  | r | r   }
\hline
&  {\bf Twitter}  & {\bf Tumblr} &  {\bf Yelp}\\
\hline
Number of unique users 			& \small   $458,117$ 	 &  \small $612,425$ & \small  $29,293$ \\ 
Number of unique items 			&  \small $1.6M$  	&  \small $1.4M$ &  \small $18,149$ \\ 
Number of unique tags 			&  \small $550,157$ 	 &  \small $2.3M$ &  \small $177,286$  \\ 
Number of triples 				&  \small $13.9M$ 	 & \small  $11.3M$  &  \small $30.3M$ \\ 
Avg number of tags per item  	&  \small $8.4$ 	& \small  $7.9$ &  \small  $685.7$  \\ 
Avg tag length				& $13.1$ & $13.0$ & $6.5$ \\
\hline
\end{tabular}
\end{center}
\vspace{-5mm}
\caption{Statistics on the datasets we used in our experiments.
}
\label{tab:StatDataset}
\end{table}
\vspace{-3mm}

\paragraph*{Twitter}
We used a collection of tweets extracted during Aug. $2012$. 
As described in Section~\ref{sec:model}, we  see  each tweet and its re-tweet instances as one item, and the authors of the tweets/re-tweets as its taggers. We include both the text and the hashtags as tags.



\vspace{-3mm}
\paragraph*{Tumblr} We extracted a collection of Tumblr posts from Oct.-Nov. $2014$, following the same interpretation on posts, taggers, and tags as in Twitter. 
Among the $6$ different types of posts within Tumblr, we selected only the \emph{default} type, which can contain text plus images. 
Moreover, in the case of Tumblr, we were able to access the follower-followee network and thus we extracted the induced follower-followee network for the selected taggers.

\vspace{-3mm}
\paragraph*{Yelp} Lastly, we considered a publicly available Yelp dataset, containing reviews for businesses and the induced follower-followee network.\footnote{http://www.yelp.com/dataset\_challenge} In this case, in order to build the triples, we considered the business (e.g., restaurant) as the item, the author of the review as the tagger, and  the keywords appearing in the review as the tags.

On Twitter and Tumblr datasets, in order to  enrich the set of keywords associated to an item, we  also expand each tag by the at most $5$ most common keywords associated with it  by a given user, i.e., by the tag-keyword co-occurrence. Finally, from the resulting sets of triples, we removed those corresponding to  (i) items that were not tagged by at least two users, or (ii) users who did not tag at least two items. 

To complete the data setting for our algorithm, we then constructed  the  user-to-user weighted networks that are exploited in the social-aware search.  For this,  we first used the underlying social network (when available). Specifically, for each user pair in Tumblr or Yelp, we computed the Dice coefficient corresponding to the common neighbors in the social network. To also study situation when such a network may not be  available (as for Twitter), exploiting a thematic proximity instead of a social one,  we built two other kinds of user similarity  networks,  based on the Dice coefficient over either (i) the item-tag pairs of the two users, or (ii) the tags of the two users. We considered the filtering of ``noise'' links, weighted below a given threshold (as discussed in Section~\ref{sec:results}). 



\vspace{-1mm}
\subsection{Experimental results: effectiveness}
\label{sec:results}
We present in this section the results we obtained in our experiments for effectiveness, or ``prediction  power'', with the purpose of validating the underlying as-you-type query model and the feasibility of our approach. In this framework, for all the  data configurations  we considered for effectiveness purposes,  we  imposed  wall-clock time thresholds of  $50ms$ per keystroke, which we see as appropriate for an interactive search experience.  

To measure effectiveness, we followed an assumption used in recent literature,   e.g. in~\cite{Vahabi2012,Maniu13-1}, namely that  a user is likely to find  his items --  belonging to him or re-published by him -- more interesting than random items from other users. 
For testing effectiveness, we randomly select triples ($u$,$i$,$t$)  from each dataset.  For each selected triple, we consider $u$ as the seeker and $t$ as the keyword issued by this user. The aim is to ``get back''  item $i$ through  search. The as-you-type scenario is simulated by considering that the user issues $t$ one letter at a time. Note that an item may be retrieved back only if at least one user connected to the seeker tagged it. 
 We picked randomly $800$ such triples (we denote this selection as the set $D$),  for tags having  at least three  letters. For each individual measurement, we gave as input a triple \emph{(user, item, tag)} to be tested (after  removing it from the dataset), and then we observed the  ranking of \emph{item} when \emph{user} issues a query that is a prefix of \emph{tag}.

Note that we tested effectiveness using single-word search for Twitter and Tumblr. On the contrary, for Yelp, due to its distinct features of having many triples per user, we did two-word search: given a query $q = (w_1, w_2)$, we first filtered items tagged by $w_1$, we then processed the remaining triples with query $w_2$ in the same manner as we did for Twitter and Tumblr.

We define the precision $P@k$ for our selected set $D$ as
$$P@k = \frac{\#\{triple \mid ranking < k, triple \in D\}}{\#D}\vspace{-1mm}$$
Since this precision can be seen as  a function of the main parameters of our system, our  goal was to understand how it is influenced by these parameters. We describe below the different parameters  we took into account here.
\begin{itemize}
\vspace{-1mm}
  \item $l$,   length of the prefix in the query (number of characters).
  \vspace{-1mm}
  \item $\theta$, the  threshold used to filter similarity links keeping only those having a score above.
  \vspace{-1mm}
  \item $\alpha$, the social bias ($\alpha = 0$ for exclusively social score, $\alpha = 1$ for exclusively textual score).
  \vspace{-1mm}
  \item $\eta_i(u)$,  the number of items tagged by user $u$, a user activeness indicator (for simplicity, hereafter referred to as $\eta_i$).
  \vspace{-1mm}
  \item $\eta_u(i)$, number of users who tagged item $i$, an item popularity indicator ($\eta_u$).
  \vspace{-1mm}
\end{itemize}

\begin{figure}[t!]
\centering
\includegraphics[width=0.225\textwidth]{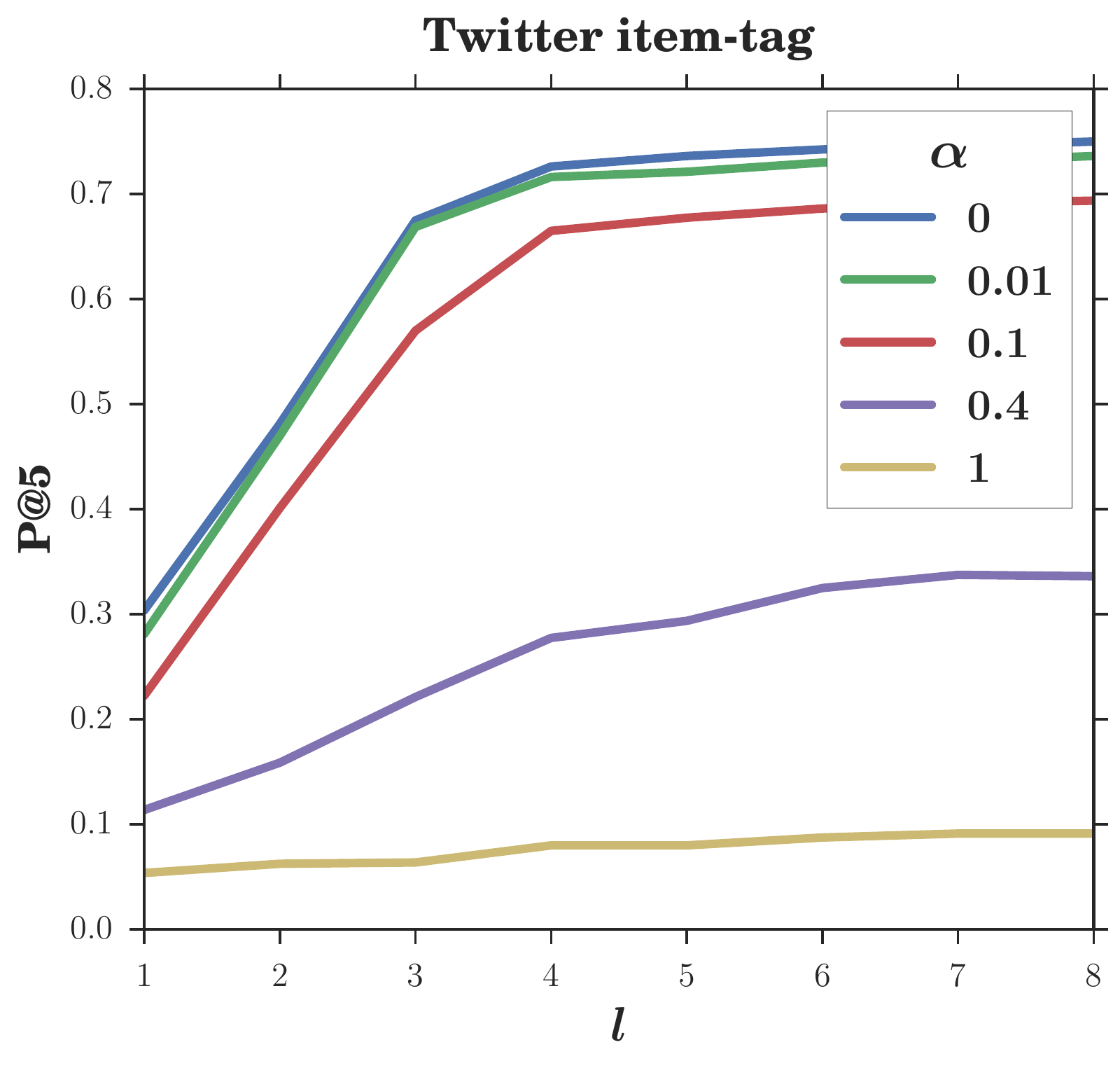}
\includegraphics[width=0.225\textwidth]{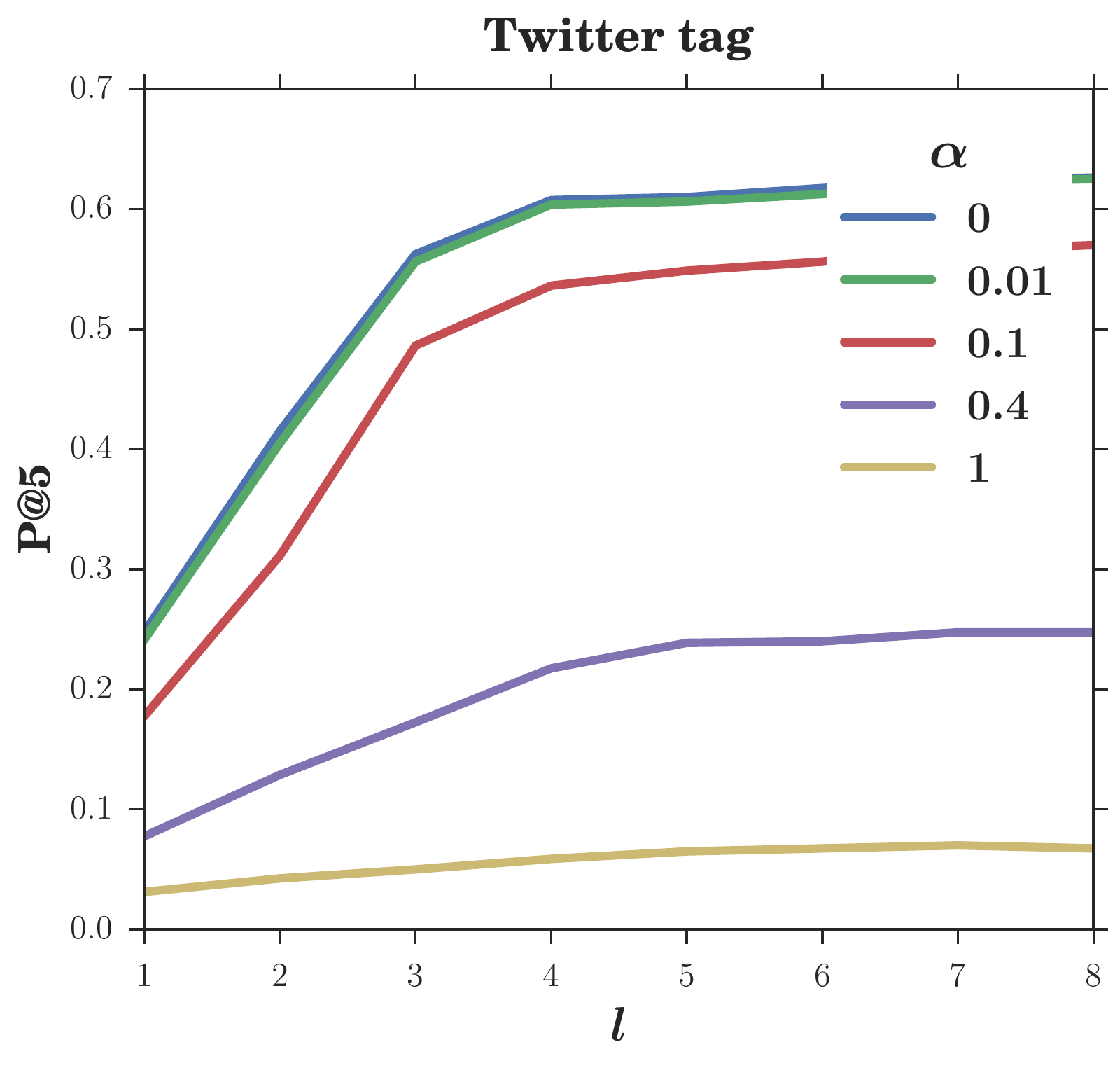}

\includegraphics[width=0.225\textwidth]{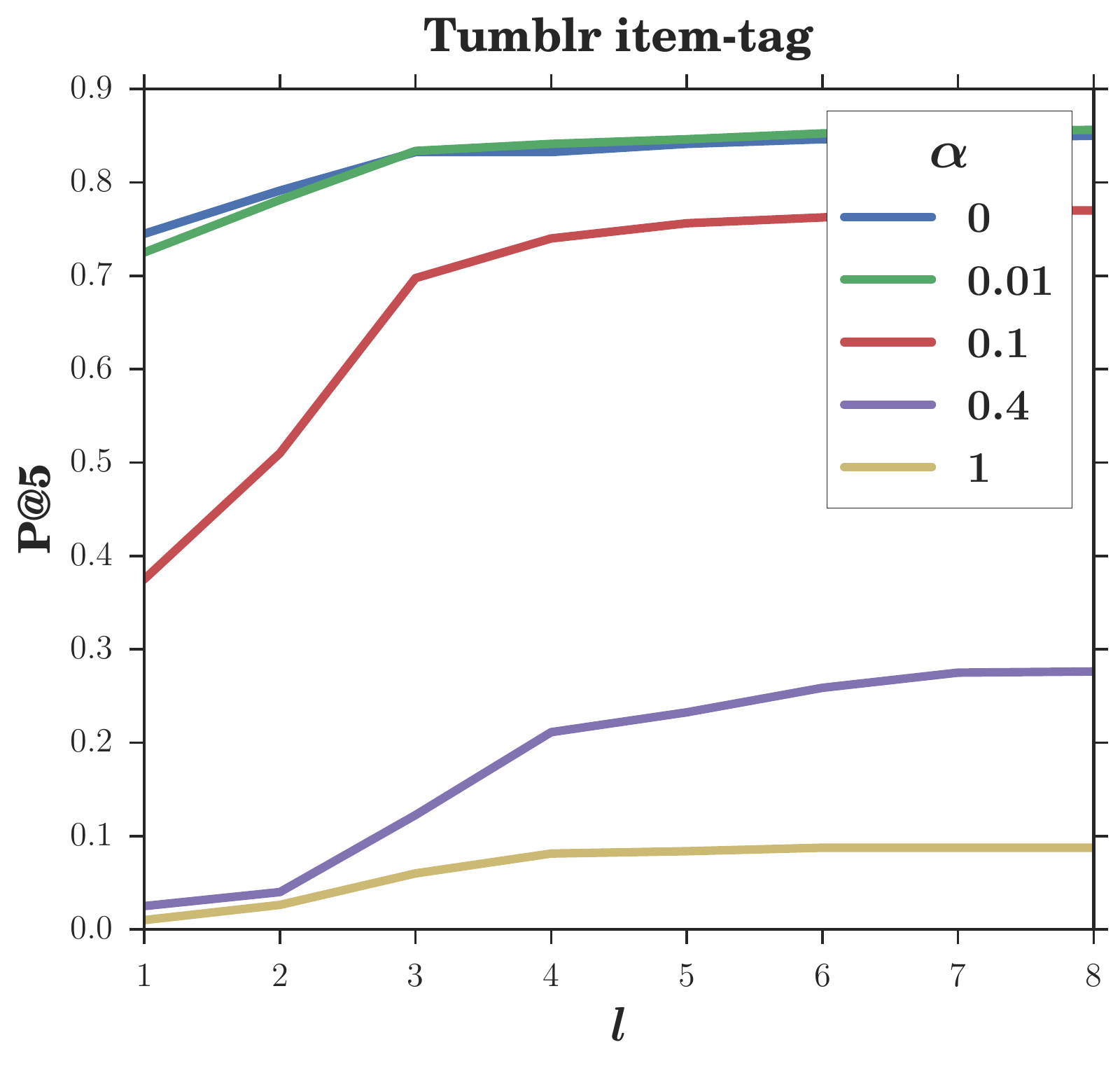}
\includegraphics[width=0.225\textwidth]{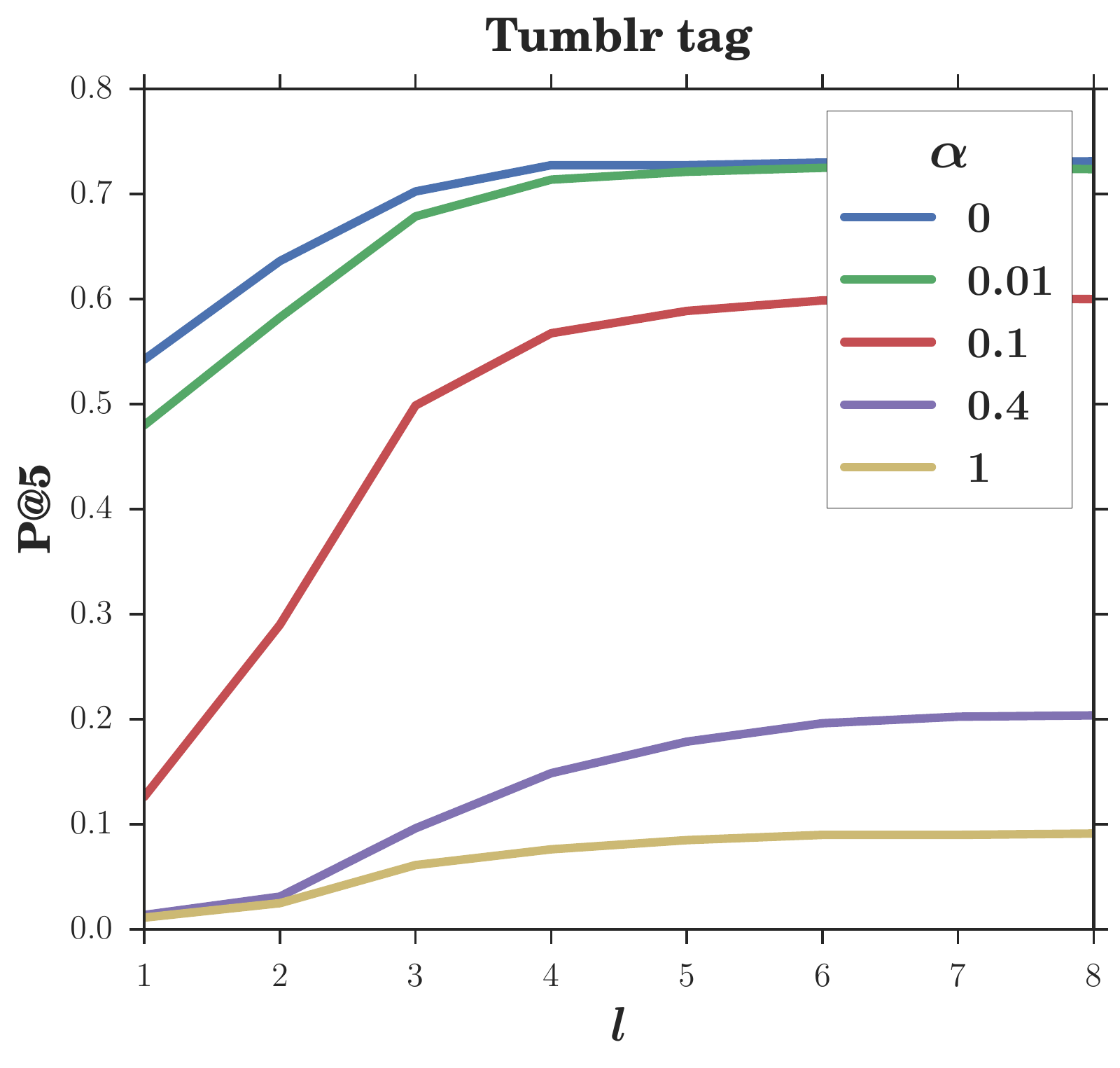}

\includegraphics[width=0.225\textwidth]{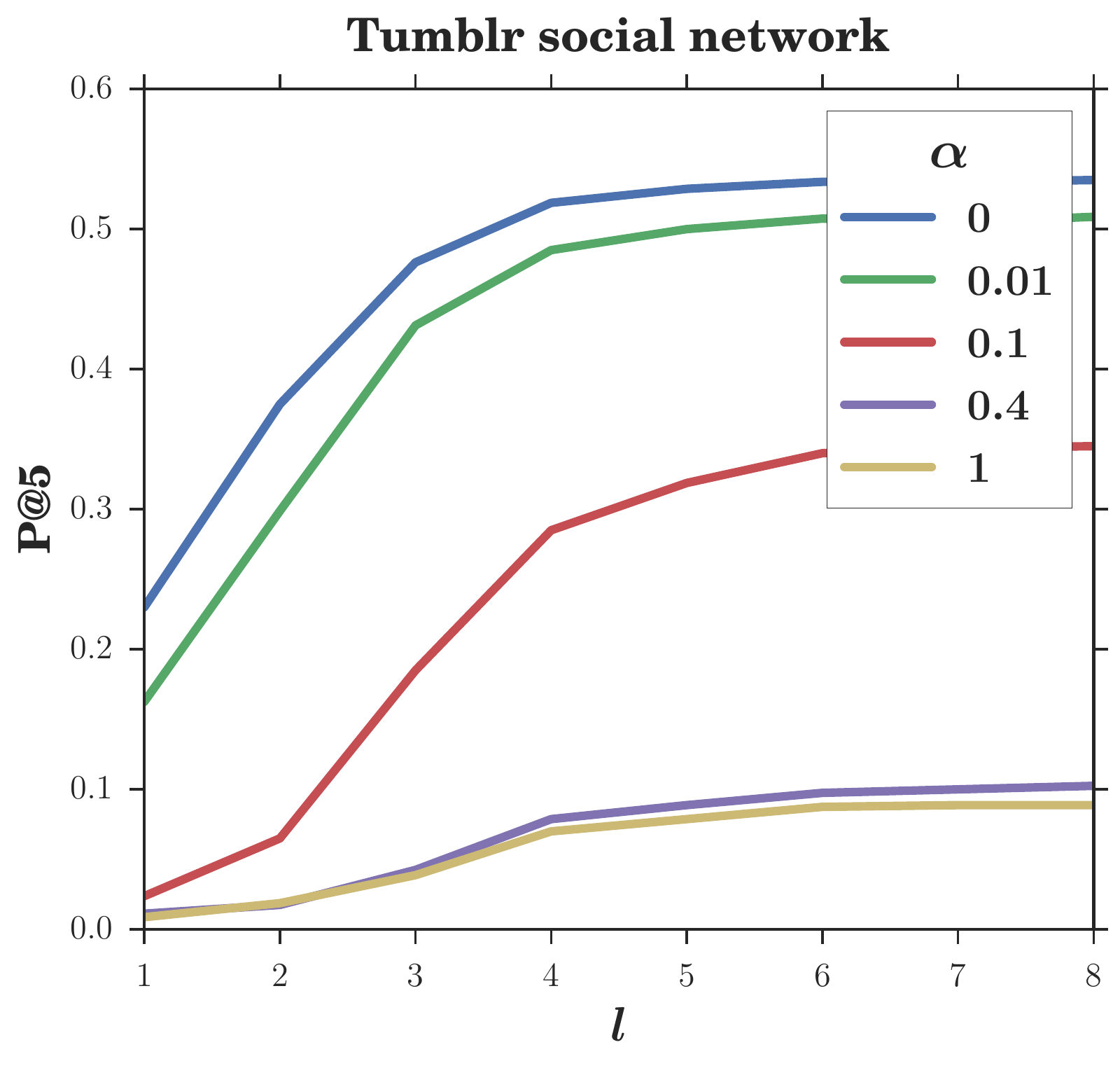}
\includegraphics[width=0.225\textwidth]{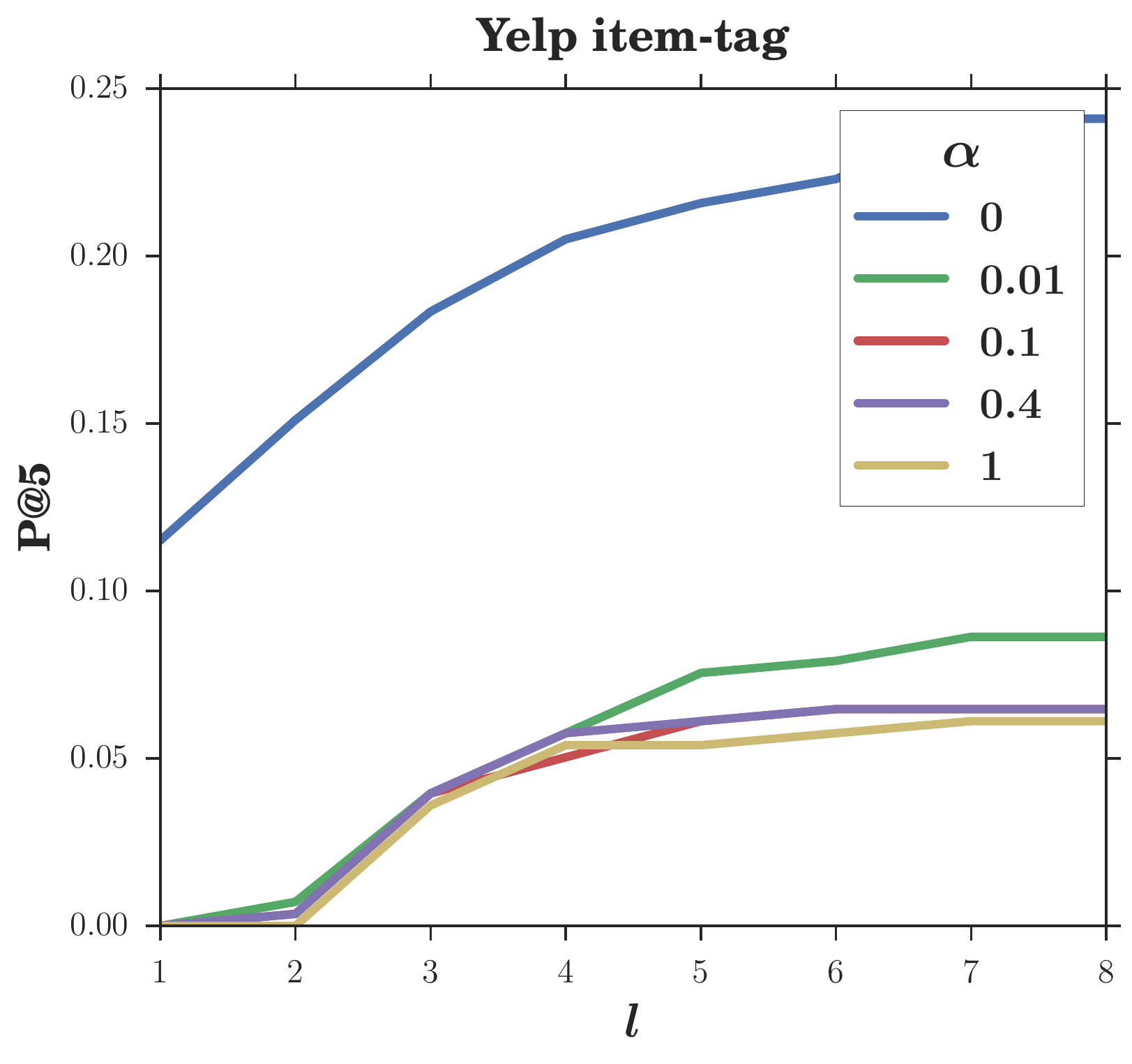}

\includegraphics[width=0.225\textwidth]{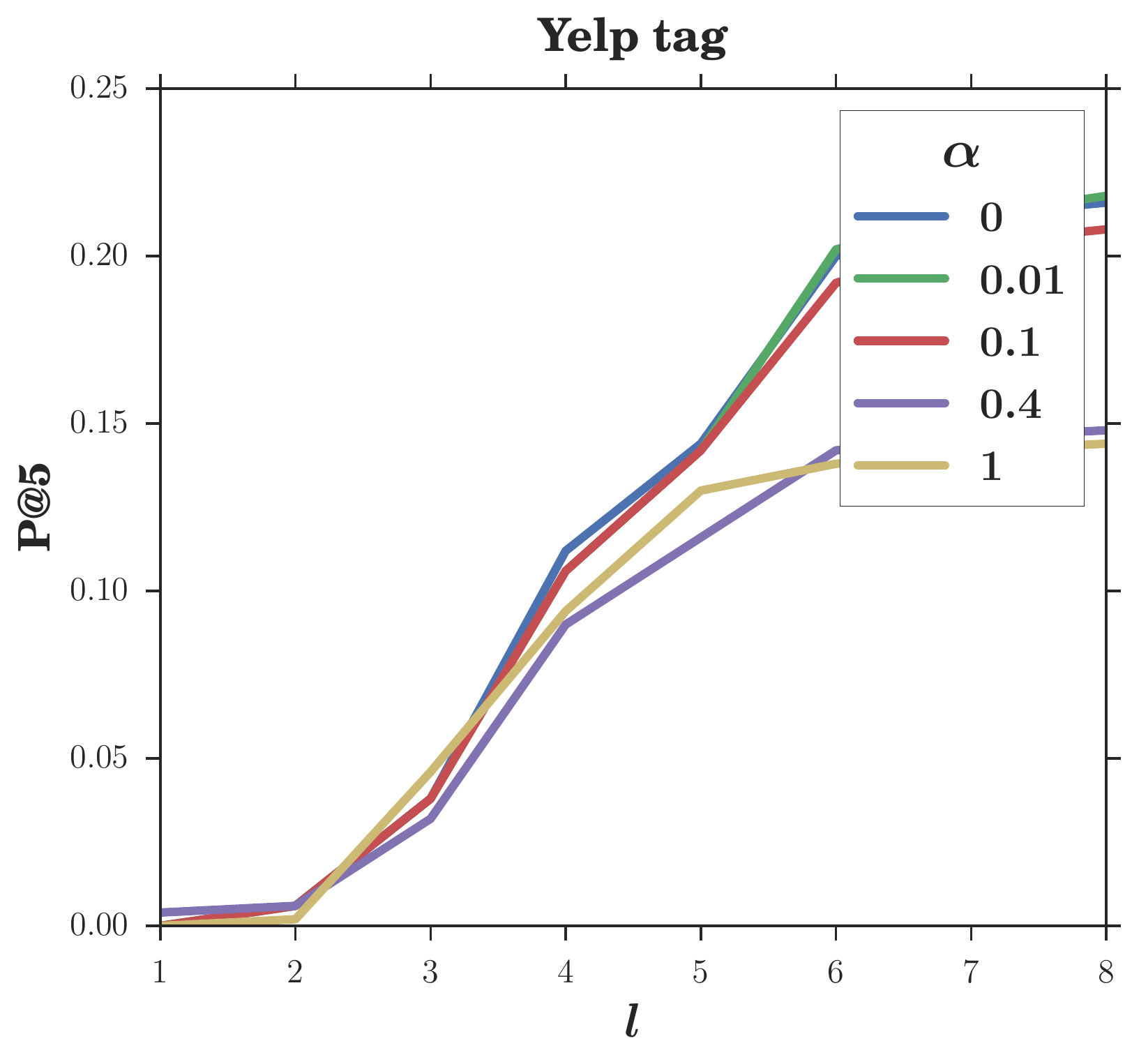}
\includegraphics[width=0.225\textwidth]{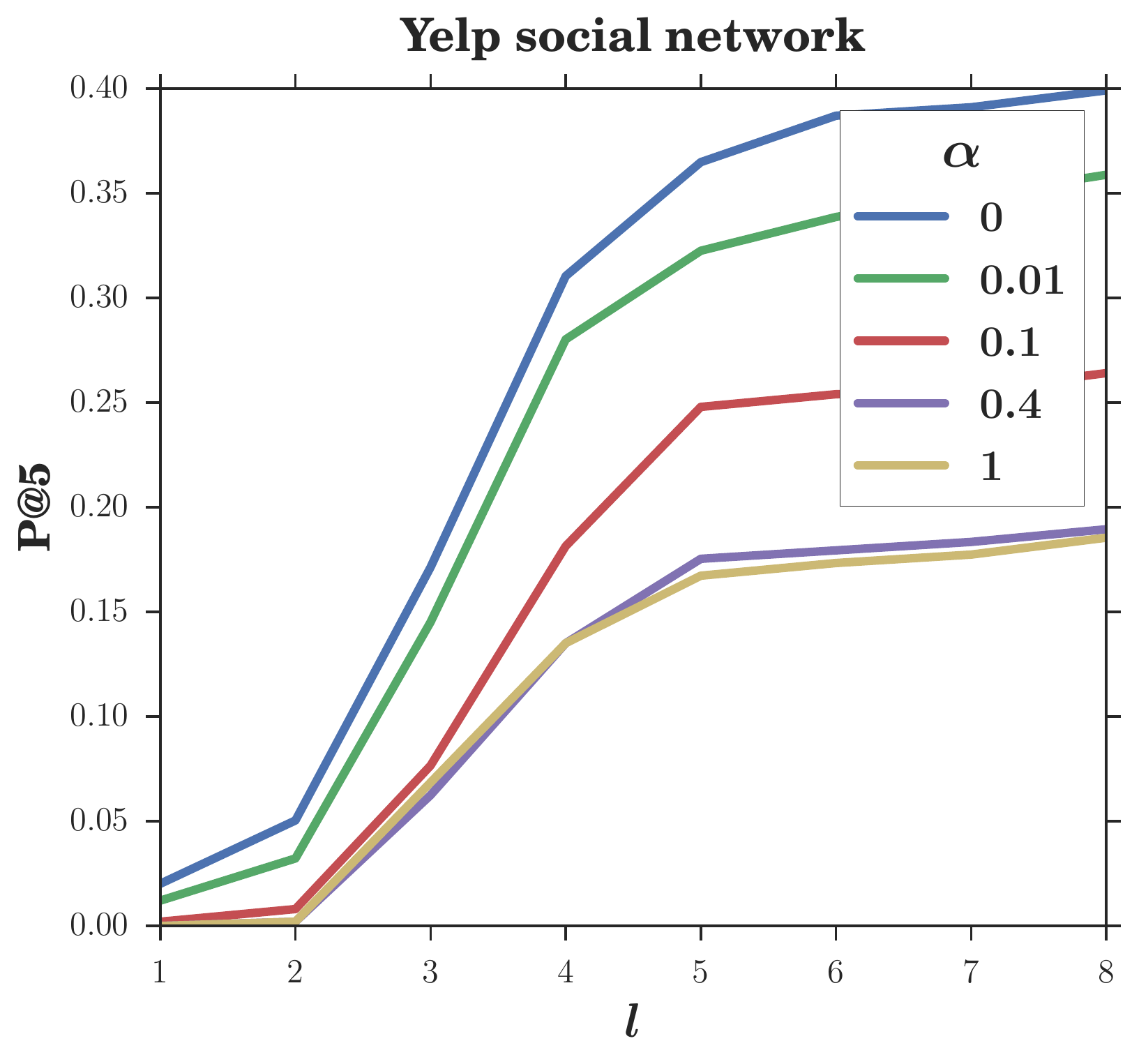}
\vspace{-5mm}
\caption{\small \label{fig:alpha2} Impact of $\alpha$ on precision.}
\end{figure}

\begin{figure}[t!]
\centering


\includegraphics[width=0.225\textwidth]{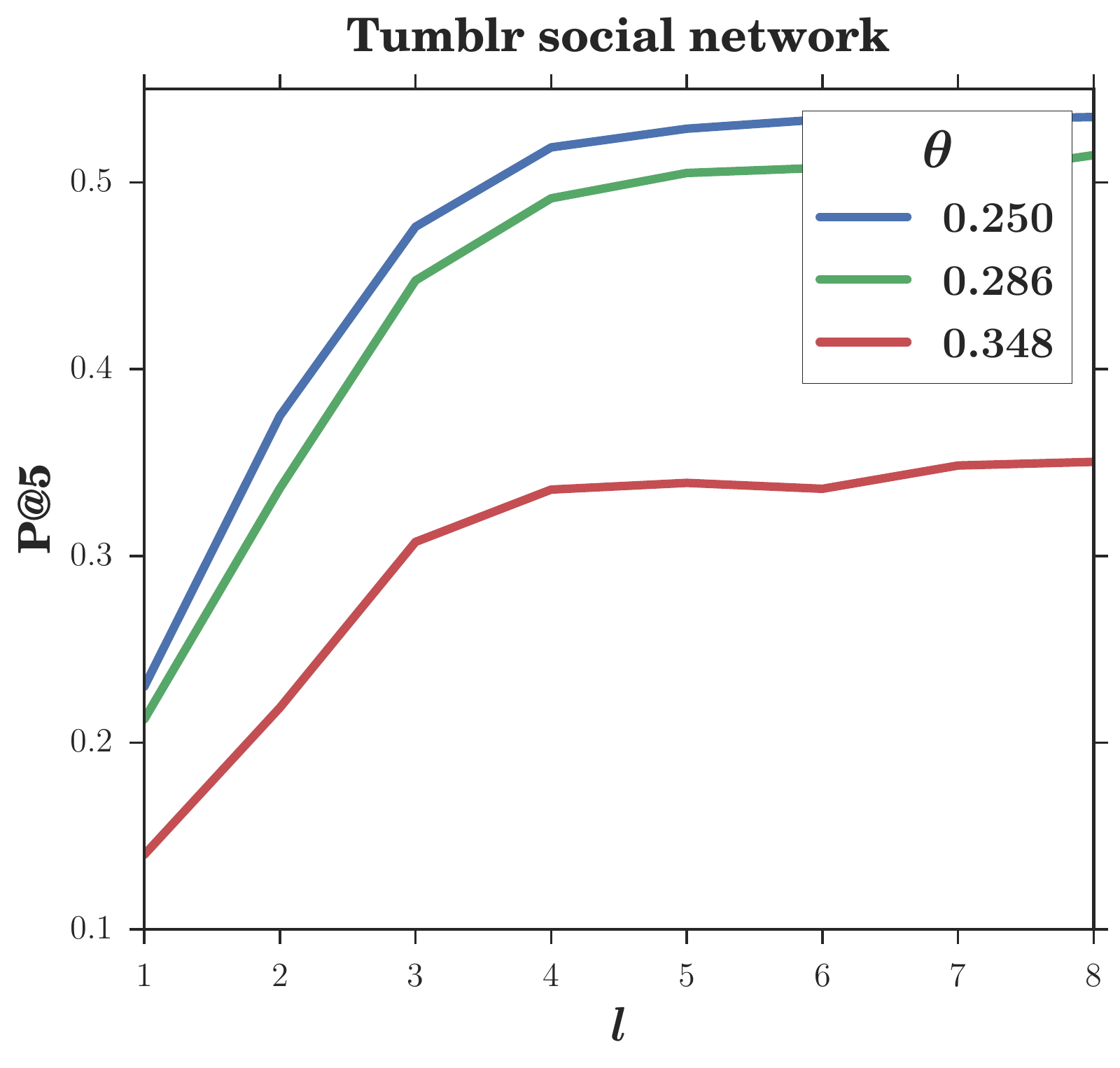}
\includegraphics[width=0.225\textwidth]{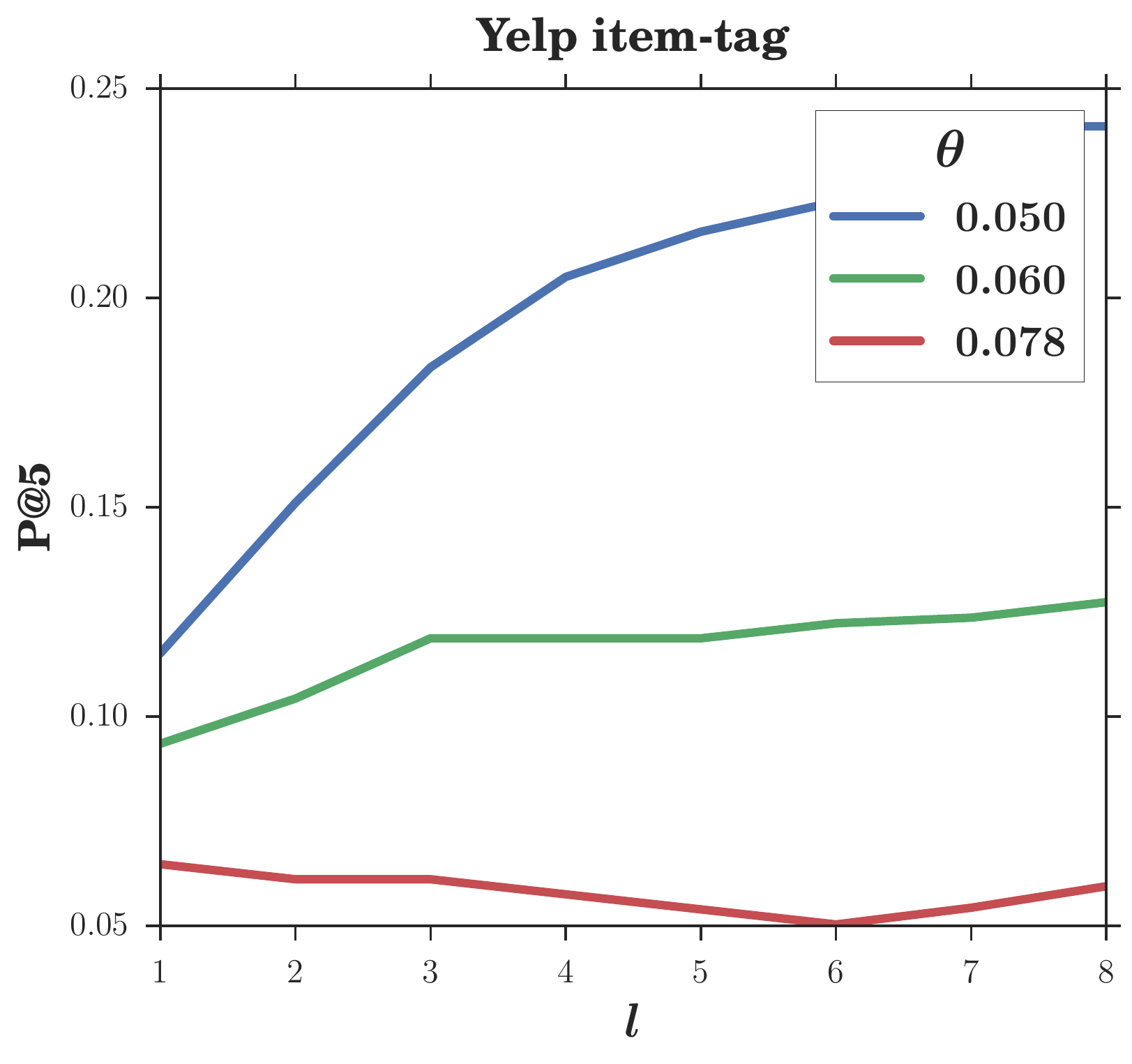}

\includegraphics[width=0.225\textwidth]{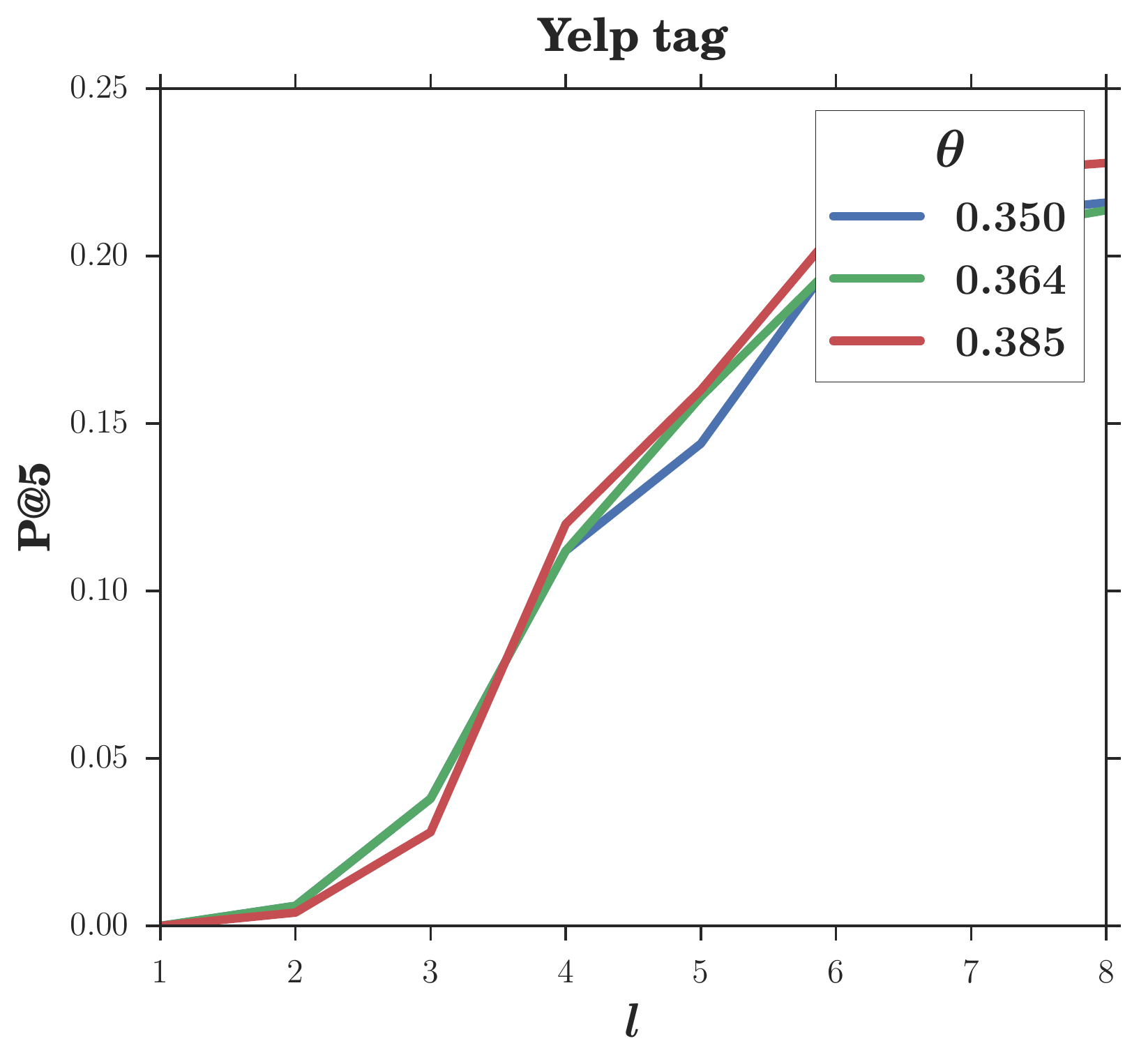}
\includegraphics[width=0.225\textwidth]{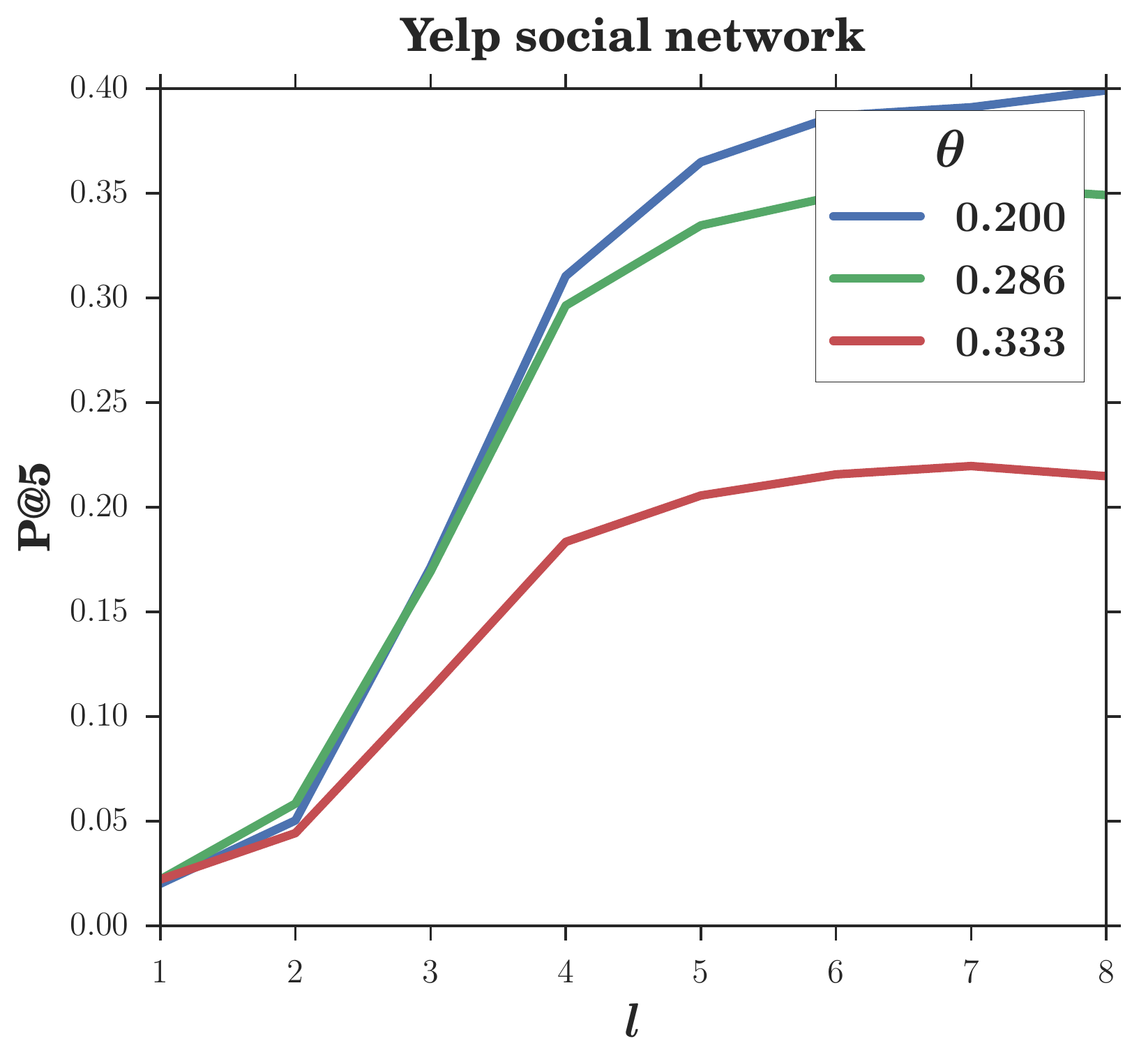}

\vspace{-5mm}
\caption{\small \label{fig:theta}Impact of $\theta$ on precision.}
\end{figure}

\noindent We present next  the results we obtained for this experiment. 
(For space reasons, we only report here on $P@5$, but we performed test with $P@1$ and $P@20$ as well, which showcase similar evolution and improvement ratios,  in the case of the latter, most often reaching precision levels of around 0.8-0.9.) 
 When parameters are not variables of a figure, they take the following default values: $\alpha = 0$ (fully social bias), $\theta$ is assigned the lowest value of the tested dataset, $\eta_i$ and $\eta_u$ are associated to active users and popular items ($\eta_i \geq 3$ and $\eta_u \geq 10$).

\paragraph*{Impact of $\alpha$} As shown in Figure~\ref{fig:alpha2}, $\alpha$ can have a major impact on precision. With a fully social bias ($\alpha = 0$), we obtained the best results for the three datasets and all the available similarity networks. Moreover, typing new characters to complete the prefix increases the precision. However, the evolution for $\alpha = 0$ can be quite slow, with the Tumblr or Yelp item-tag similarity network for witness. In this case, one likely reason is that these networks are quite rich in information, and the neighbors of the seeker are very likely to have the searched item, with the right tag,  due to the way this network was built. 
This can also  explain why the precision for the item-tag networks is higher in the case of Tumblr than those for tag and social similarity networks. The precision for the social similarity network is the lowest for Tumblr, while in the case of Yelp dataset the best results are obtained using the social network.
Indeed, the tag and item-tag networks were built based on the same content we were testing on, whereas the social similarity network only uses the links between users to infer distances between them. Yelp exhibits lower precision levels overall, unsurprisingly, since it is a much denser dataset (number of triples per user). 

Interestingly and supporting our thesis for social bias, we obtain good precisions levels with such networks of similarity in social links (the highest in the case of Yelp). For example, in the case of Tumblr, we can reach  $P@5$ of around $0.82$ for the item-tag similarity network,  $0.7$ for the tag one, and still $0.5$ for the social one. This indicates that we can indeed find relevant information using a content-agnostic network using \algoname. Importantly,  it also indicates that we can always search with the same social similarity network, even when the content evolves rather rapidly, with the same precision guarantees.


\paragraph*{Impact of $\theta$} In Figure \ref{fig:theta}, we can observe the impact of $\theta$ on the quality of results.  We mention that the two highest $\theta$ values lead to 33\% and 66\% cuts on the total number of edges obtained with the lowest $\theta$ value. Unsurprisingly, removing connections between users decreases the precision. When using the similarity network filtered by the lowest $\theta$ value, the seeker is almost always connected to the network's largest connected component, and we can visit many users to retrieve back the targeted item. With higher $\theta$ values, the connectivity for certain seekers we tested with is broken, making some of the tested items unreachable.

\begin{figure}
\centering


\includegraphics[width=0.225\textwidth]{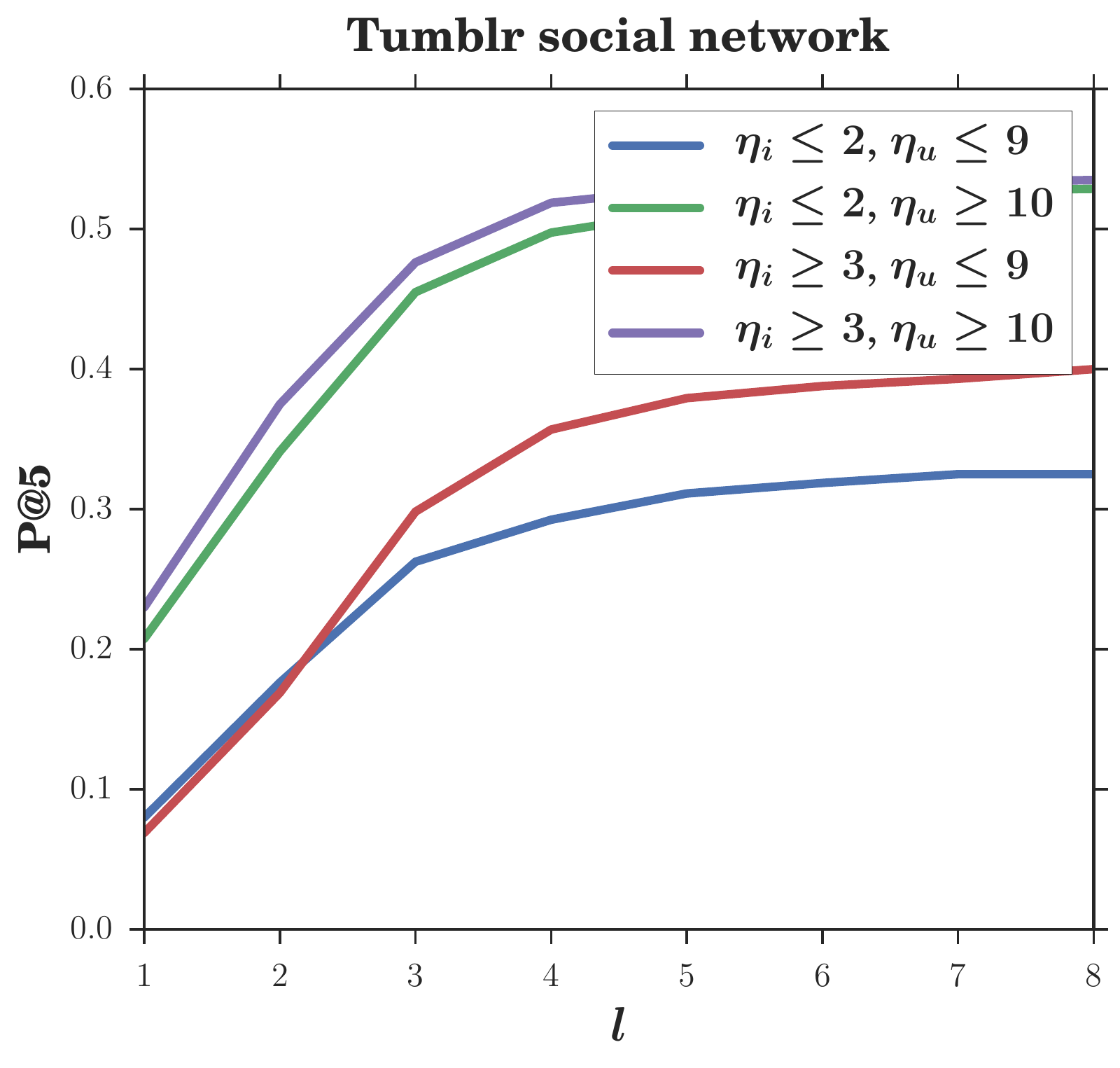}
\includegraphics[width=0.225\textwidth]{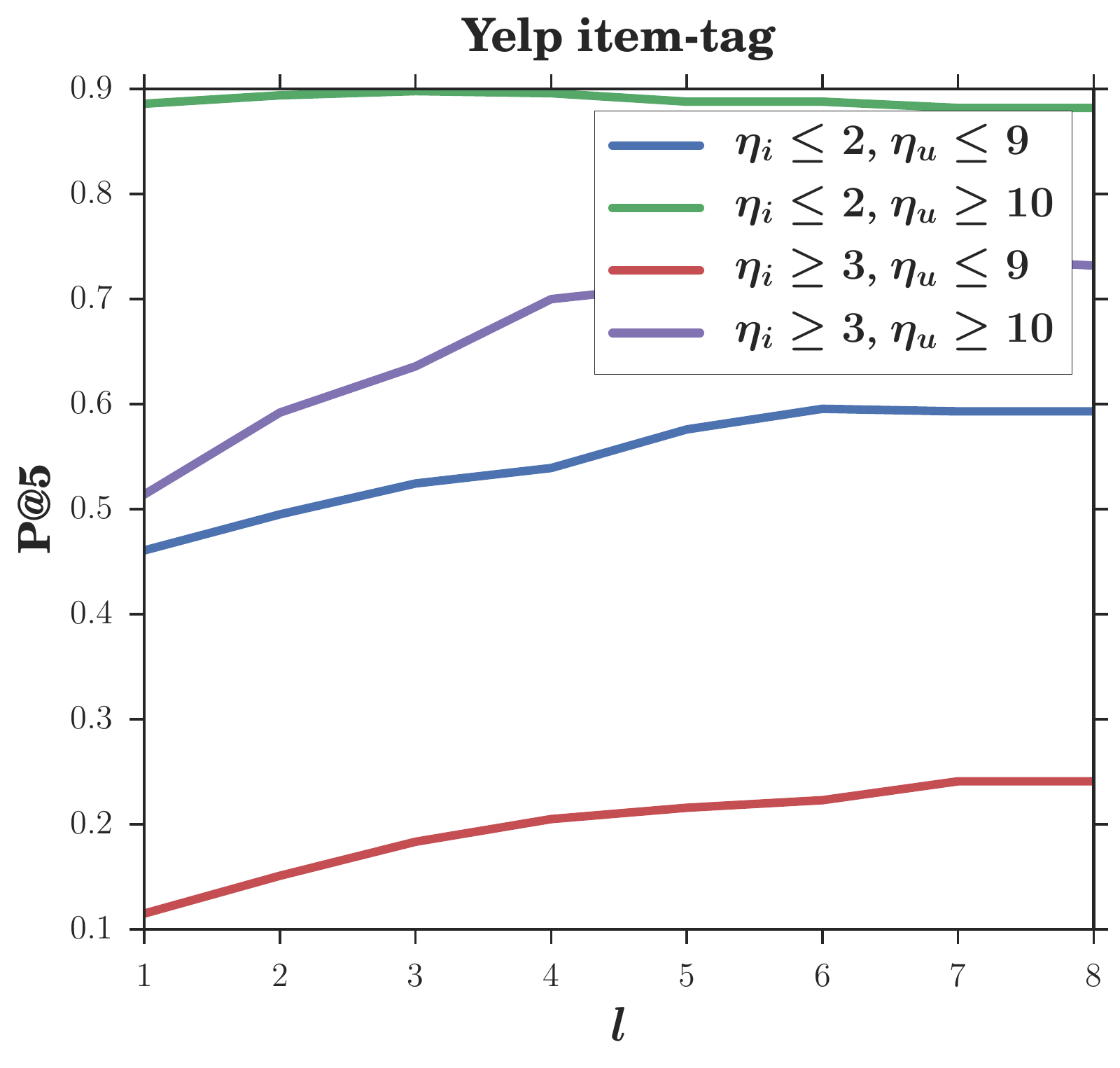}

\includegraphics[width=0.225\textwidth]{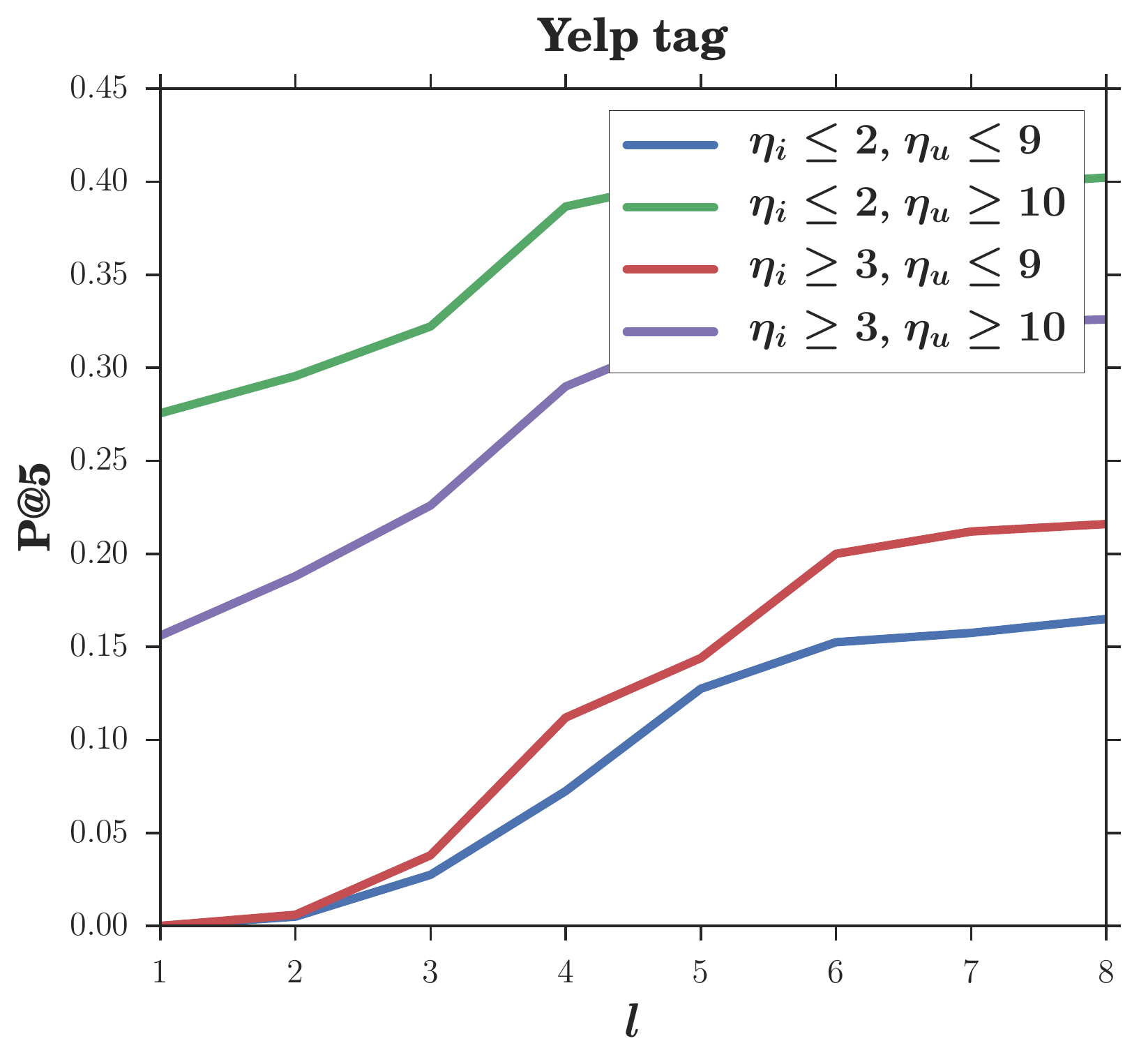}
\includegraphics[width=0.225\textwidth]{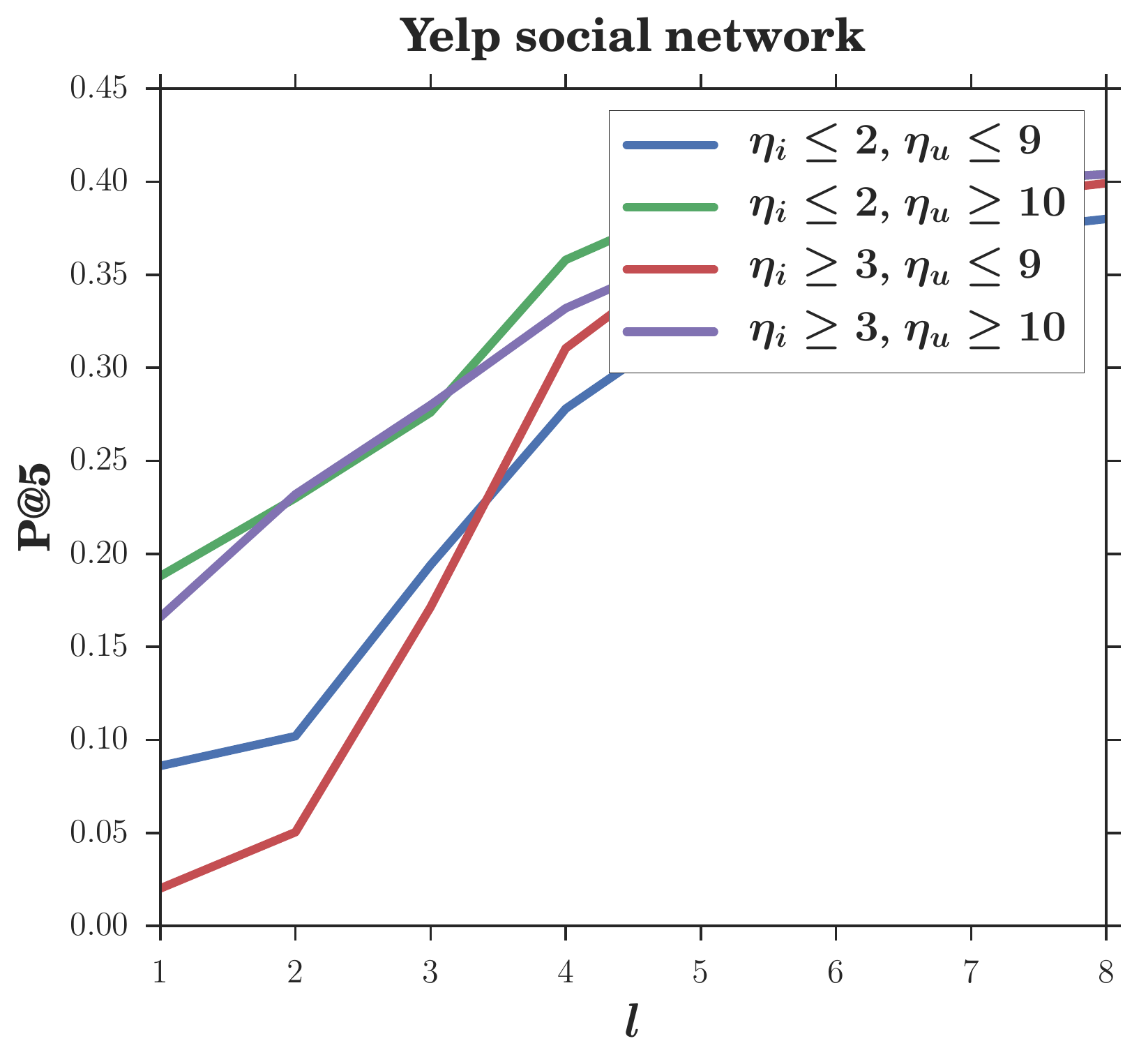}

\vspace{-5mm}
\caption{\small \label{fig:filtering} Precision for various types of users and items.}
\end{figure}

\paragraph*{Impact of popularity / activeness}
We show in Figure \ref{fig:filtering}  the effects of item popularity and user activity for Yelp and Tumblr. For all similarity networks, the precision is better for popular items (high $\eta_u$). This is to be expected, as a popular item is more likely to be found when visiting the graph, as it is expected that it will score high since it has many taggers. Along with item popularity, we can observe that user activeness has a different effect in both  content-based and the social similarity networks. Active users yield a better precision score when similarity comes from social links, whereas it is the opposite with content-based similarity networks. Reasonably, retrieving back an item for a non-active seeker in a content-based network is easier since his similarity with  neighbours is stronger (Dice coefficient computed on less content).

\subsection{\hspace{-3.3mm}Experimental results:~efficiency\&scalability}
\label{sec:resultsefficency}

\begin{figure}[t]
\centering
\includegraphics[width=0.235\textwidth]{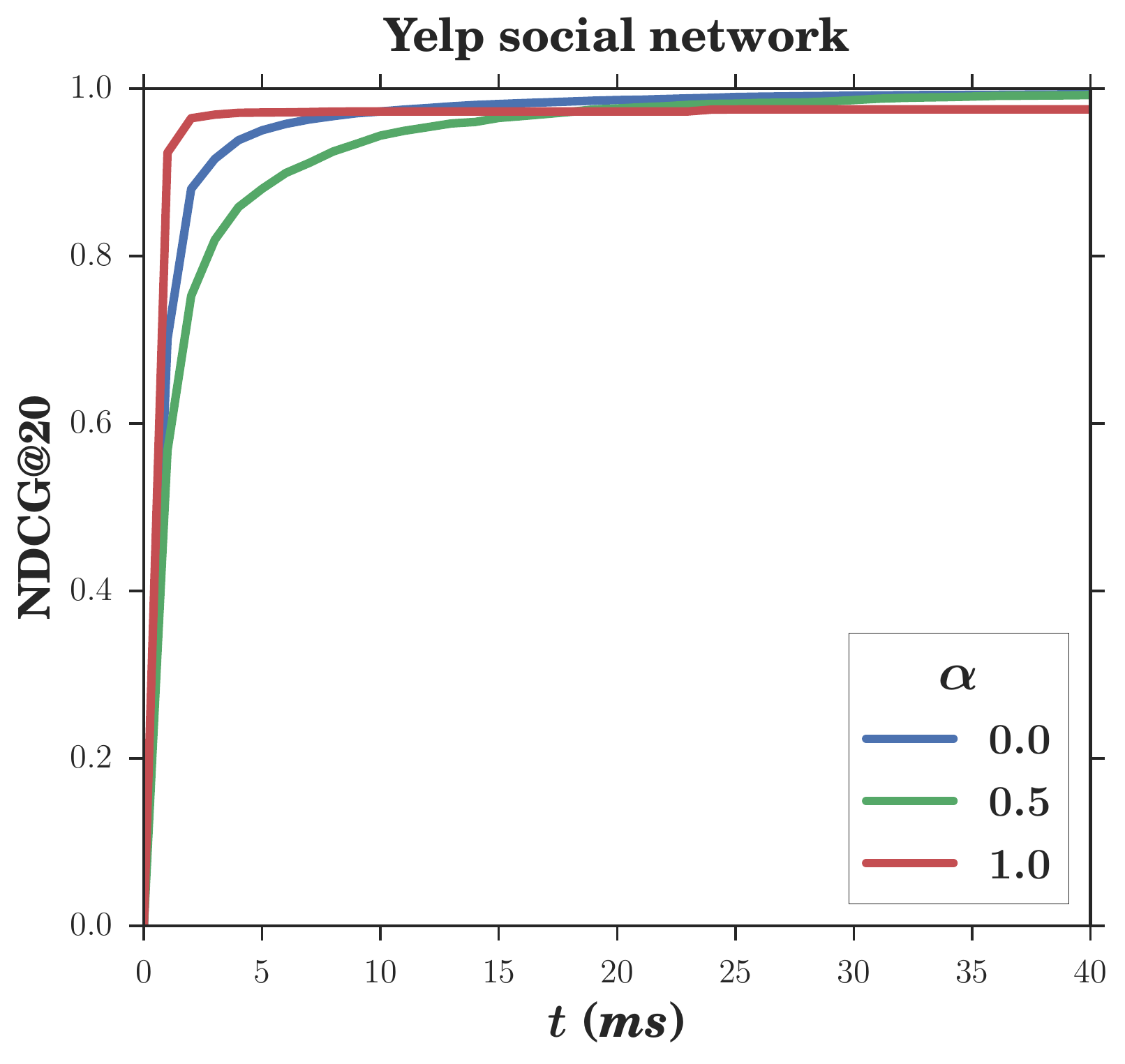}
\includegraphics[width=0.235\textwidth]{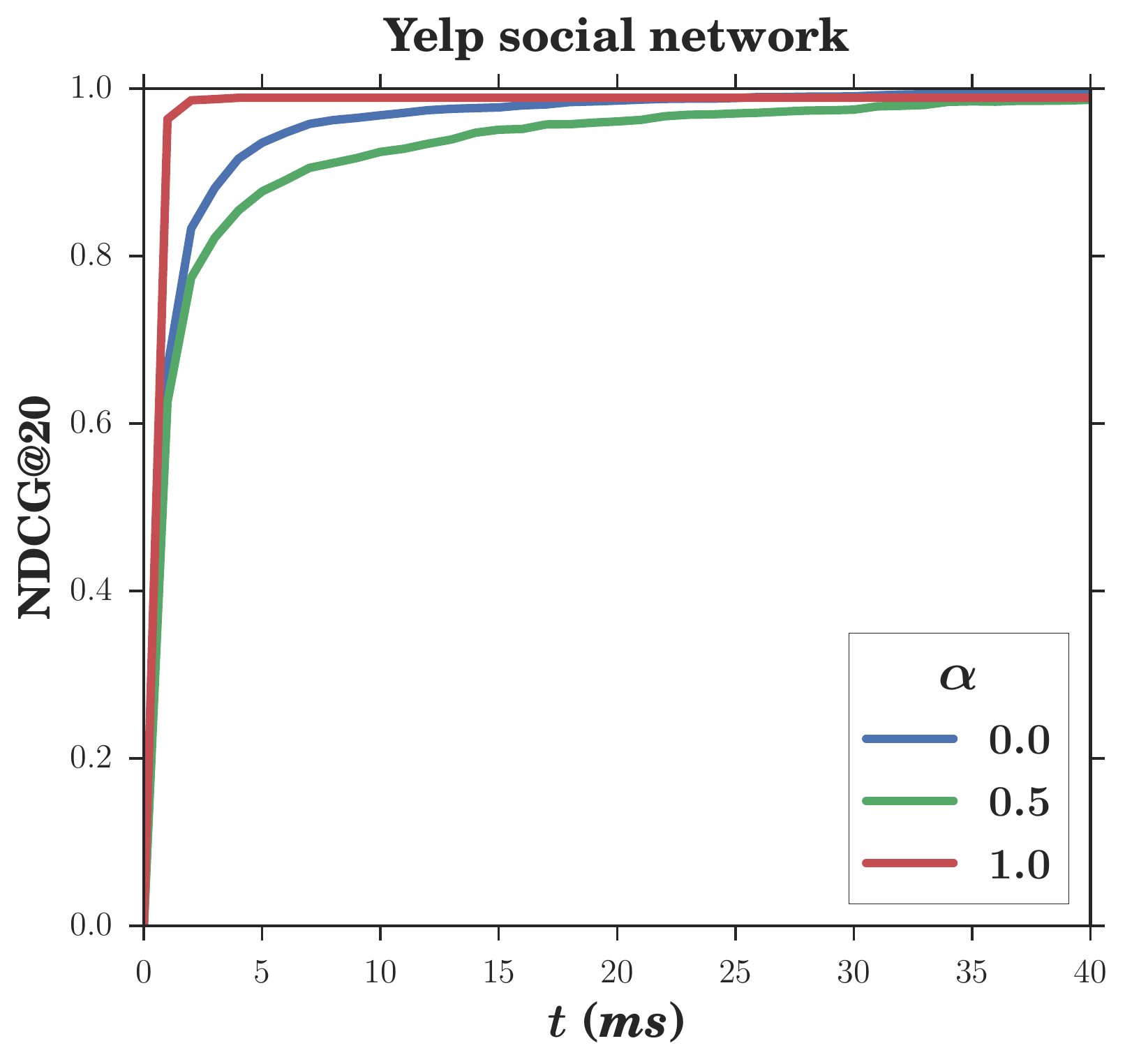}
\vspace{-8.5mm}
\caption{\small \label{fig:NDCGAlpha} Impact of $\alpha$ on NDCG vs time for random search (left) and personal search (right).}
\end{figure}

In Figure~\ref{fig:NDCGAlpha}, we display the evolution of NDCG@20 vs. time, for the  densest dataset (Yelp), for different $\alpha$ values (where $\alpha$ is normalized to have similar social and textual scores in average). The NDCG is computed w.r.t. the exact top-$k$,  that would be obtained running the algorithm on the entire similarity graph. This measure is an important indicator for the feasibility of social-aware as-you-type search, illustrating the accuracy levels reached under "typing latency'', even when the termination conditions are not met. In this plot, we fixed the prefix length size to $l=4$. The left plot is when a user searches with a random tag (not necessarily used by her previously), while the right plot follows the same selection methodology as in Section~\ref{sec:results}. Importantly, with $\alpha$ corresponding to an exclusively social or textual relevance, we reach the exact top-$k$ faster than when combining these two contributions ($\alpha = 0.5$). Note also that this trend holds even when the user searches with random tags. 

\begin{figure}
\centering
\vspace{-4mm}
\includegraphics[width=0.235\textwidth]{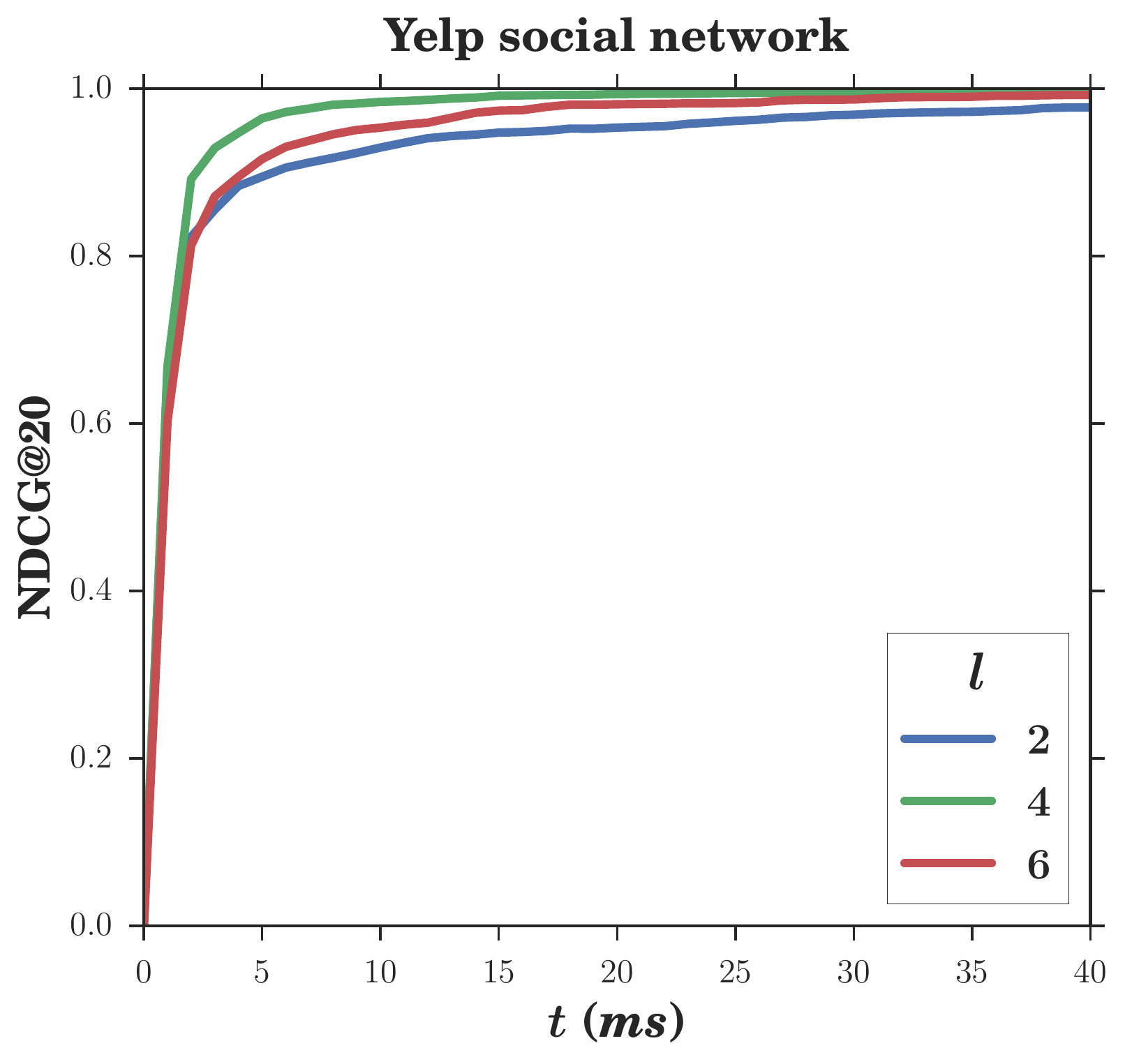}
\includegraphics[width=0.235\textwidth]{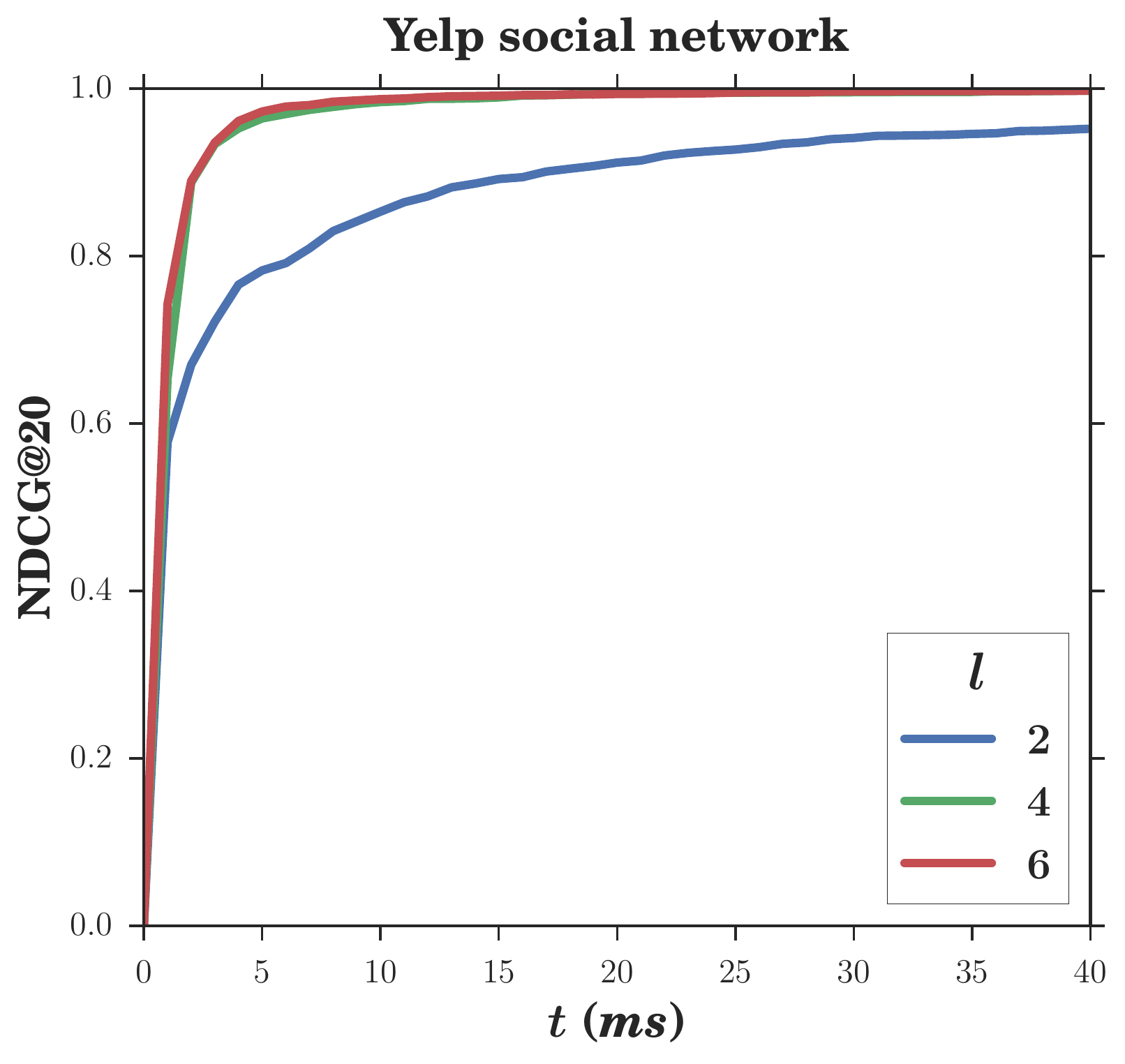}
\vspace{-8.5mm}
\caption{\small \label{fig:NDCGtimeL} Impact of $l$ on NDCG vs time for random search (left) and personal search (right).}
\end{figure}

In Figure~\ref{fig:NDCGtimeL}, similarly to the previous case,  we show the evolution of NDCG@20 vs. time in Yelp,  for different prefix lengths. 
(the left plot is for random tags).  Results shows that with lower values of $l$ we need more time to identify the right top-$k$. The reason is that shorter prefixes can have many potential (matching) items, therefore the item discrimination process evolves more slowly.

\begin{figure}[t!]
\centering
\vspace{-4mm}
\includegraphics[width=0.235\textwidth]{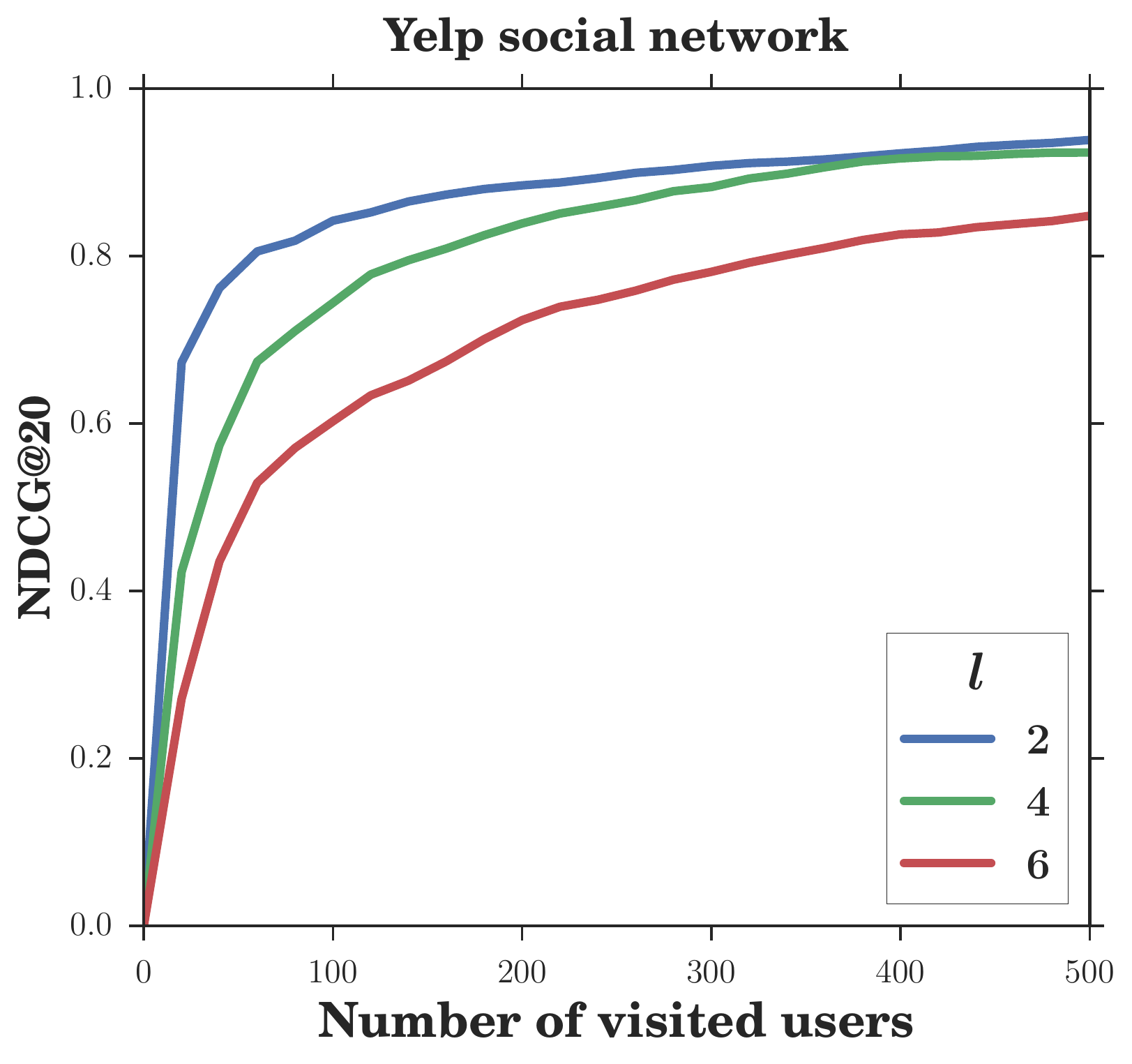}
\includegraphics[width=0.235\textwidth]{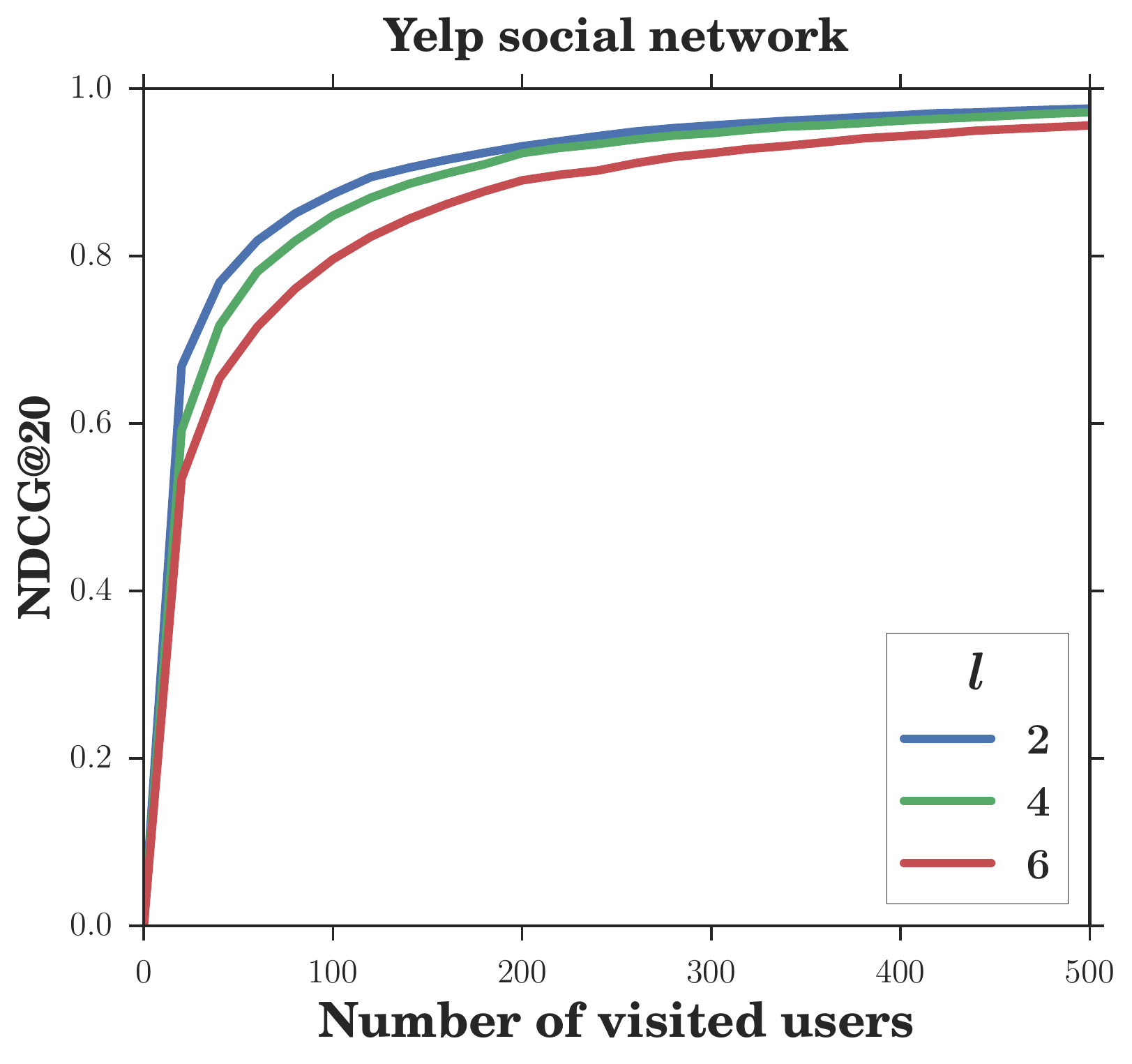}
\vspace{-8.5mm}
\caption{\small \label{fig:NDCGFigure} Impact of $l$ on NDCG vs number of visited users for random search (left) and personal search (right). }
\end{figure}

In Figure~\ref{fig:NDCGFigure} we show the evolution of NDCG@20 when visiting a fixed number number of users. We show results for $l=2,4,6$. As expected, the more users we visit the higher NDCG we reach. For longer prefixes, it is necessary to visit more users. For instance, when $l=6$, after visiting $500$ users, we reach an NDCG of $0.8$ while for $l=2$ the NDCG after $500$ visits is $0.9$.


\begin{figure}
\centering
\includegraphics[width=0.252\textwidth]{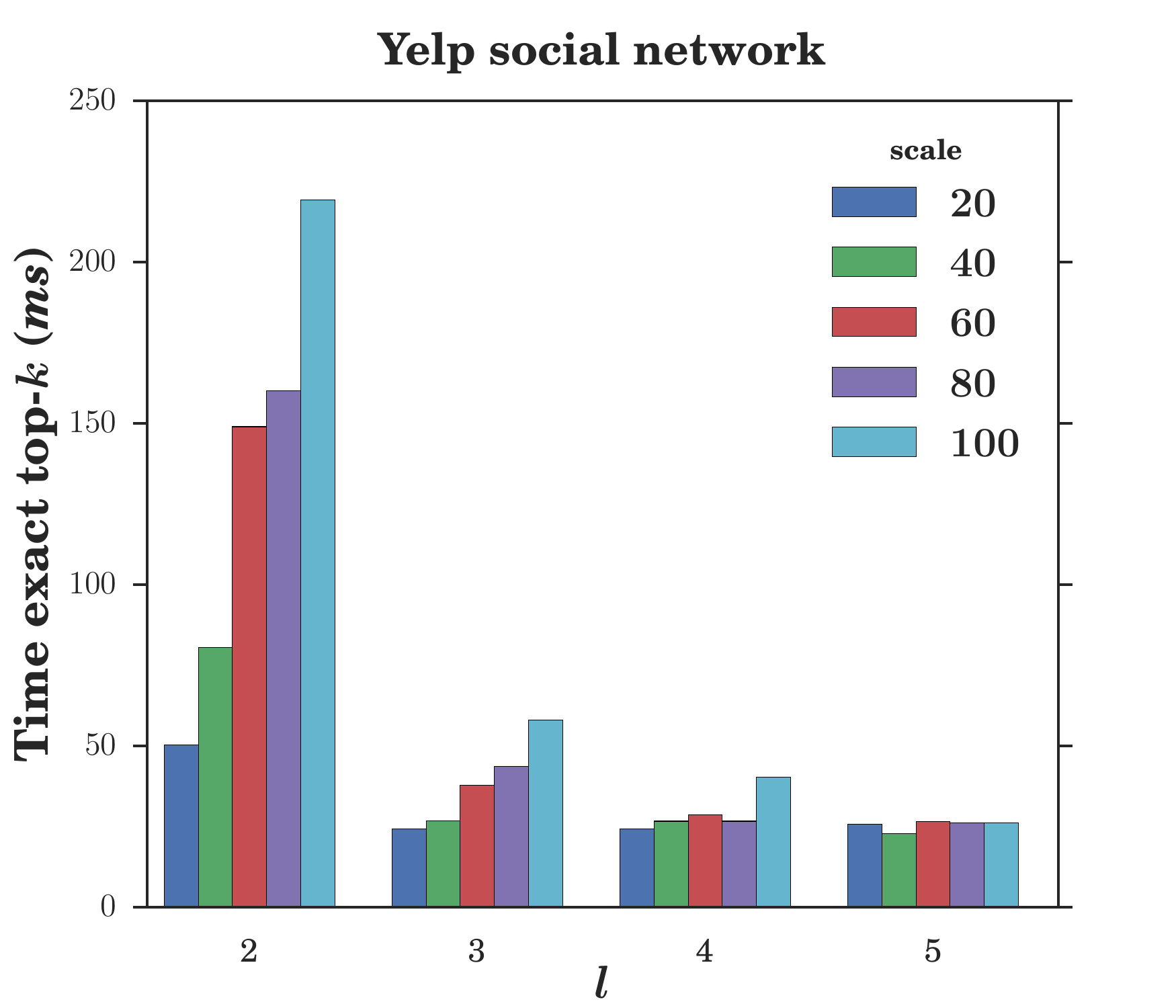}
\vspace{-5mm}
\caption{\small \label{fig:scalability}  Time to exact top-k for different dataset sizes.}
\end{figure}

Finally, in the experiment illustrated in Figure~\ref{fig:scalability} we observed the time to reach the exact top-$k$ for different dataset sizes. For that, we partitioned the Yelp triples sorted by time into five consecutive  ($20\%$) chunks. For each dataset we perform searches using prefixes of $l=2,3,4,5$. While the time to reach the exact top-$k$ increases with bigger datasets and shorter prefixes, the algorithm scales adequately when $l$ is more than $2$. For instance, for $l=3$, the time to reach the result over the complete dataset is just twice the time when considering only $20\%$ of this dataset.

\paragraph*{Main-memory vs. secondary memory considerations} We emphasize here that we  performed our experiments in an all-in-memory setting, for datasets of medium size (tens of millions of tagging triples), in which the advantages of our approach may not be entirely observed. In  practice, in  real, large-scale applications such as Tumblr, one can no longer assume a direct and cheap access to p-spaces and inverted lists, even though some data dimensions such as the user network and the top levels of  \indexname\ -- e.g., the trie layer and possibly prefixes of the inverted lists -- could still reside in main memory.  In practice, with each visited user,  the search might require a random access for her personal space, hence the interest for the sequential, user-at-a-time approach. Even when p-spaces may reside on disk, our last experiment shows that by retrieving a small number of them, less than 100, we can reach good precision levels; depending on disk latency, serving results in, for example, under $100ms$ seems within reach. One way to further alleviate such costs may  be to cluster users having similar proximity vectors, and choose the layout of p-spaces on disk based on such clusters; this is an approach we intend to evaluate in the future, at larger scale.

\section{Conclusion}
\label{sec:conclusion}

We study in this paper as-you-type  top-$k$ retrieval in social tagging applications, under a network-aware query model by which information produced by users who are closer to the seeker can be given more weight. We formalize this problem and we describe the \algoname\ algorithm to solve it, based on  a novel trie data structure, \indexname, allowing ranked access over inverted lists. In several application scenarios, we perform extensive experiments for efficiency, effectiveness, and scalability, validating our techniques and the underlying query model. As a measure of efficiency, since as-accurate-as-possible answers  must be provided while the query is being typed, we investigate how precision evolves with time and, in particular, under what circumstances acceptable precision levels are met within reasonable as-you-type latency (e.g., less than $50ms$).  Also, as  a measure of effectiveness, we analyse thoroughly the ``prediction power'' of the results produced by \algoname.

We see many promising directions for improving the \algoname\ algorithm.  First, for  optimising query execution over the \indexname\  index structure,  we intend to study how  \indexname\  can be  enriched with  certain pre-computed unions of inverted lists (materialised virtual lists). 
Assuming a fixed memory budget, this would be done for chosen nodes (prefixes) in the trie, in order to speed-up the sorted access time, 
leading to a memory-time tradeoff.  While similar in spirit to the pre-computation of virtual lists of~\cite{Li:2012}, a major difference for our setting is that we can rely on a materialization strategy guided by the social links and the tagging activity, instead of one guided by a known query workload. 
Also, one difficult case in our as-you-type scenario is the one in which $t_r$ is the initial character, following a number of already completed query terms. One possible direction for optimisation in \textsc{TOPKS-ASYT} is to avoid revisiting users, by  recording the  accessed p-spaces for future reference. In short, within the memory budget, a na\"ive solution would be to keep these p-spaces as such (one per user). However, in order to speed-up the ranked retrieval, a more promising solution is to organise the p-spaces in a completion trie as well, which would allow us to access their entries by order of relevance.   

\section*{Acknowledgement}
\vspace{-0.06mm}
This work was partially supported by the French research project ALICIA (ANR-13-CORD-0020) and by the EU research project LEADS (ICT-318809).

\nocite{*}
\balance
\small
\bibliographystyle{abbrv} 
\bibliography{paper-short}



\end{document}